\documentclass[aps,prc,superscriptaddress,floatfix]{revtex4-1}

\usepackage[english]{babel}											
\usepackage{amsmath}										
\usepackage{amssymb}                    
\usepackage{amsfonts}
\usepackage{graphicx,xcolor,textpos}
\usepackage{epstopdf}
\usepackage{placeins}
\usepackage{dcolumn}

\begin{document}
\title{Empirical description of beta-delayed fission partial half-lives}
 \author{L.~Ghys}
\email[]{lars.ghys@fys.kuleuven.be}
\affiliation{KU Leuven, Instituut voor Kern- en Stralingsfysica, 3001 Leuven, Belgium}
\affiliation{Belgian Nuclear Research Center SCK$\bullet$CEN, Boeretang 200, B-2400 Mol, Belgium}

\author{A.N.~Andreyev}
\affiliation{Department of Physics, University of York, York, YO10 5DD, United Kingdom}
\affiliation{Advanced Science Research Center, Japan Atomic Energy Agency, Tokai-Mura, Naka-gun, Ibaraki, 319-1195, Japan}

\author{S.~Antalic}
\affiliation{Departement of Nuclear Physics and Biophysics, Comenius University, 84248 Bratislava, Slovakia}

\author{M.~Huyse}
\author{P.~Van Duppen}
\affiliation{KU Leuven, Instituut voor Kern- en Stralingsfysica, 3001 Leuven, Belgium}
\date{\today}

\begin{abstract}
\begin{description}
\item[Background] The process of $\beta$-delayed fission ($\beta$DF) provides a versatile tool to study low-energy fission in nuclei far away from the $\beta$-stability line, especially for nuclei which do not fission spontaneously.
\item[Purpose] The aim of this paper is to investigate systematic trends in $\beta$DF partial half-lives.
\item[Method] A semi-phenomenological framework was developed to systematically account for the behavior of $\beta$DF partial half-lives.
\item[Results] The $\beta$DF partial half-life appears to exponentially depend on the difference between the $Q$ value for $\beta$ decay of the parent nucleus and the fission-barrier energy of the daughter (after $\beta$ decay) product. Such dependence was found to arise naturally from some simple theoretical considerations.
\item[Conclusions] This systematic trend was confirmed for experimental $\beta$DF partial half-lives spanning over 7 orders of magnitudes when using fission barriers calculated from either the Thomas-Fermi or the liquid-drop fission model. The same dependence was also observed, although less pronounced, when comparing to fission barriers from the finite-range liquid-drop model or the Thomas-Fermi plus Strutinsky Integral method.
\end{description}
\end{abstract}

\maketitle

\section{Introduction}
\begin{figure}
\includegraphics[width=0.5\columnwidth]{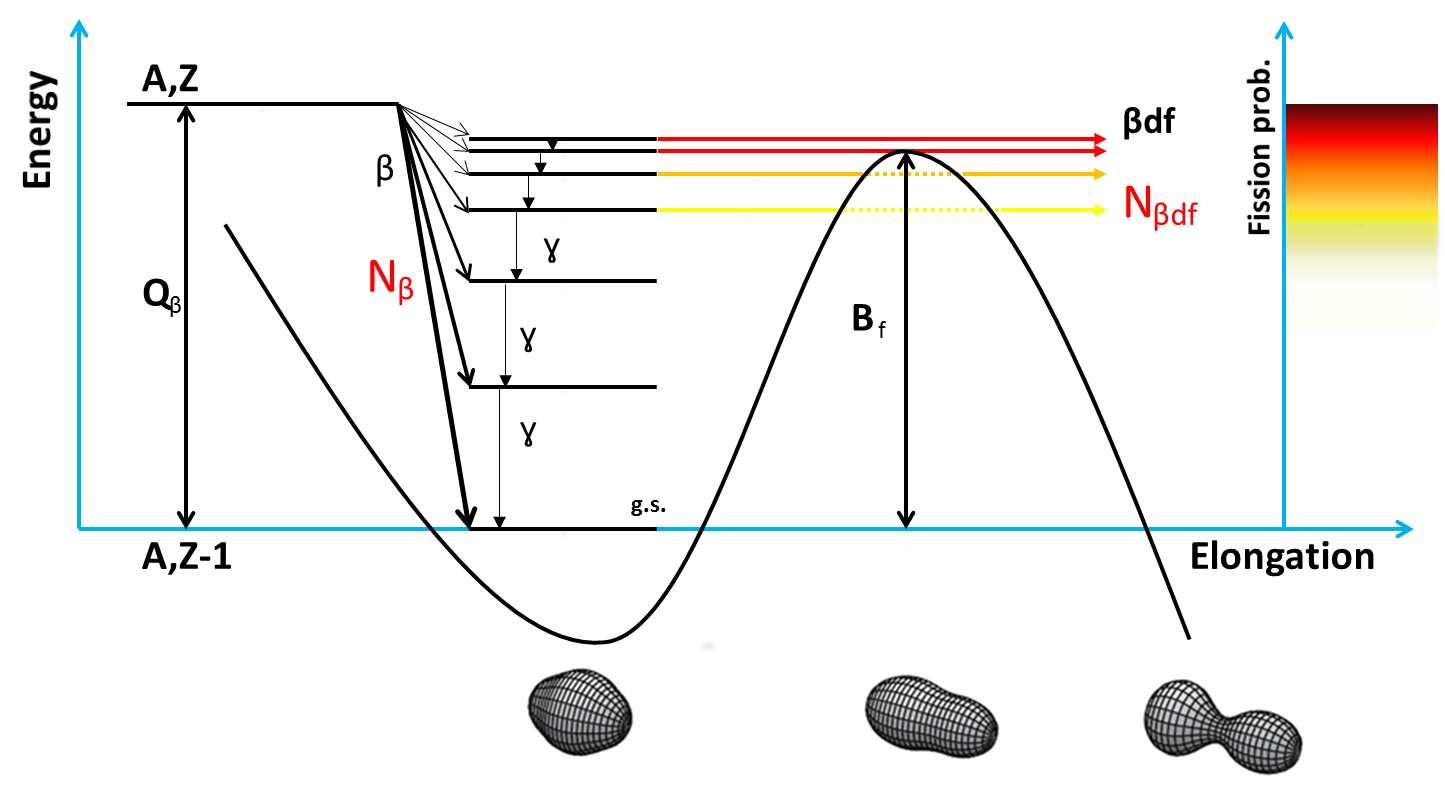}
\caption{(Color online) Schematic representation of the $\beta$DF process on the neutron-deficient side of the nuclear chart. The $Q_{EC}$ value of the parent (A,Z) nucleus is indicated, while the curved line shows the potential energy of the daughter (A,Z-1) nucleus with respect to nuclear elongation, displaying also the fission barrier $B_f$.  The color code on the right-hand side represents the probability for excited states, with excitation energies close to $B_f$, to undergo fission; the darker colors correspond to higher probabilities.}
\label{fig:bdfSchem}
\end{figure}

$\beta$-delayed fission ($\beta$DF) is a two-step process whereby the fissioning nucleus could be created in an excited state after $\beta$ decay of a precursor. Since the excitation energy of the fissioning daughter product is limited by the $Q_{\beta}$ value for $\beta$ decay of the parent, $\beta$DF provides a unique tool to study low-energy fission of nuclei far from stability, especially for those not fissioning spontaneously. Figure \ref{fig:bdfSchem} provides a schematic representation of this process, for nuclides on the neutron-deficient side of the nuclear chart. Recent experiments at ISOLDE-CERN \cite{Andreyev2010,Elseviers2013,Liberati2013,Ghys2014} and SHIP-GSI \cite{Andreyev2013, Lane2013} have studied this exotic decay mode in several short-lived neutron-deficient isotopes in the lead region. The fission-fragment mass and energy distributions resulting from $\beta$DF have established a new region of asymmetric fission around $^{178,180}\mathrm{Hg}$ \cite{Andreyev2010,Liberati2013} and indicated multimodal fission in $^{194,196}$Po and $^{202}$Rn \cite{Ghys2014}. A recent review of the $\beta$DF process is given in \cite{Andreyev2013a}, in which a total of 27 $\beta$DF cases, both on the neutron-rich and neutron-deficient sides, were summarized.

It is furthermore believed that $\beta$DF could, together with neutron-induced and spontaneous fission, influence the fission-recycling in r-process nucleosynthesis \cite{Panov2005,Petermann2012}. Therefore, a reliable prediction of the relative importance of $\beta$DF in nuclear decay, often expressed by the $\beta$DF probability $P_{\beta \mathrm{DF}}$, is needed. $P_{\beta \mathrm{DF}}$ is defined as

\begin{equation}
P_{\beta \mathrm{DF}} = \frac{N_{\beta \mathrm{DF}}}{N_{\beta}},
\label{eq:Pbdf}
\end{equation}

where $N_{\beta DF}$ and $N_{\beta}$ are respectively the number of $\beta$DF and $\beta$ decays of the precursor nucleus.
An earlier comparison of $P_{\beta \mathrm{DF}}$ data in a relatively narrow region of nuclei in the vicinity of uranium showed a simple exponential dependence with respect to $Q_{\beta}$ \cite{Shaughnessy2000,Shaughnessy2002}. It was assumed that fission-barrier heights $B_{ \rm f}$ of the daughter nuclei do not vary greatly in this region \cite{Britt1980} ($B_{ \rm f}\sim4$\,--\,6\,MeV) and thus have a smaller influence on $P_{\beta \mathrm{DF}}$ as compared to $Q_{\beta}$ values ($Q_{\beta}\sim3$\,--\,6\,MeV). In addition, these nuclei have a typical $N$/$Z$ ratio around $\sim$\,1.4\,--\,1.5, which is close to that of traditional spontaneous fission of heavy actinides.\\
The aim of this paper is to further explore such systematic features by including the newly obtained data in the neutron-deficient lead region whose $\beta$DF nuclides have significantly different N/Z ratios ($\sim$\,$1.2$\,--\,1.3), $B_{\rm f}$ ($\sim$\,7\,--\,10\,MeV) and $Q_{\beta}$ values ($\sim$\,9\,--\,11\,MeV) as compared to those in the uranium region.\\
However, from an experimental point of view, the dominant $\alpha$-branching ratio ($\gtrsim90$\,$\%$) in most $\beta$DF precursors in the neutron-deficient lead region \cite{Nubase2012} makes precise determination of $N_{\beta}$ in equation (\ref{eq:Pbdf}) difficult. Therefore, the partial $\beta$DF half-life $T_{\rm 1/2p,\beta DF}$, as proposed in \cite{Andreyev2013a}, is discussed in the present study. By analogy with other decay modes, $T_{\rm 1/2p,\beta DF}$ is defined by

\begin{equation}
T_{\rm 1/2p,\beta DF} = T_{1/2}\frac{N_{\mathrm{dec,tot}}}{N_{\beta \mathrm{DF}}},
\label{eq:Tbdf}
\end{equation}

where $T_{1/2}$ represents the total half-life and $N_{\mathrm{dec,tot}}$ the number of decayed precursor nuclei. The relation between $T_{\rm 1/2p,\beta DF}$ and $P_{\beta \mathrm{DF}}$ can be derived from equations (\ref{eq:Pbdf}) and (\ref{eq:Tbdf}) as

\begin{equation}
T_{\rm 1/2p,\beta DF} = \frac{T_{1/2}}{b_{\beta}P_{\beta \mathrm{DF}}},
\label{eq:PbdfvsTbdf}
\end{equation}

with $b_{\beta}$ denoting the $\beta$-branching ratio. If the $\alpha$-decay channel dominates, as is often the case in the neutron-deficient lead region, one can safely approximate $N_{\mathrm{dec,tot}}$ in equation (\ref{eq:Tbdf}) by the amount of $\alpha$ decays $N_{\mathrm{\alpha}}$.\\
This work shows an apparent exponential dependence of $T_{\rm 1/2p,\beta DF}$ on ($Q_{\beta}-B_{\rm f}$) for certain sets of calculated fission-barrier energies. Such relation may arise naturally by simple phenomenological approximations of the $\beta$-strength function of the precursor and the fission-decay width of excited states in the daughter nucleus. These assumptions may be justified considering the scale of the systematic trend discussed here, spanning $T_{\rm 1/2p,\beta DF}$ values over several orders of magnitude. Deviations lower than one order of magnitude are thus acceptable.

\section{Theoretical considerations}
\label{sec:Theo}
Following \cite{Gangrsky1980,Klapdor1983,Kuznetsov1999}, the expression for $P_{\beta \mathrm{DF}}$ is given by

\begin{equation}
P_{\beta \mathrm{DF}} = \frac{\int_0^{Q_{\beta}}S_{\beta}(E)F(Q_{\beta}-E)\frac{\Gamma_{\rm f}(E)}{\Gamma_{\text{tot}}(E)}dE}{\int_0^{Q_{\beta}}S_{\beta}(E)F(Q_{\beta}-E)dE},
\label{eq:PBdf}
\end{equation}

whereby the $\beta$-strength function of the parent nucleus is denoted by $S_{\beta}$ and the Fermi function by $F$. The excitation energy is here, and further, given by E. The fission and total decay widths of the daughter, after $\beta$ decay, are respectively given by $\Gamma_{\text{f}}$ and $\Gamma_{\text{tot}}$. Equation (\ref{eq:PbdfvsTbdf}) can be combined with equation (\ref{eq:PBdf}) to deduce the decay constant of $\beta$DF, defined as $\lambda_{\beta \mathrm{DF}} = \mathrm{ln}(2)/T_{\rm 1/2p,\beta DF}$, as

\begin{equation}
\lambda_{\beta \mathrm{DF}} = \int_0^{Q_{\beta}}S_{\beta}(E)F(Q_{\beta}-E)\frac{\Gamma_{\rm f}(E)}{\Gamma_{\text{tot}}(E)}dE.
\label{eq:lambdaBdf}
\end{equation}

This section will be devoted to the derivation of an analytical expression for $\lambda_{\beta \mathrm{DF}}$ by approximating $S_{\beta}$, $F$ and $\Gamma_{\rm f}/\Gamma_{\text{tot}}$. Since most of the reliable experimental data on $\beta$DF are recorded on the neutron-deficient side of the nuclear chart (see Table \ref{tab:Tbdf} and \cite{Andreyev2013a}), only EC/$\beta^+$-delayed fission will be considered here.

\subsection{Approximations}
\label{approx}

A first simplification in equation (\ref{eq:lambdaBdf}) is to approximate $S_{\beta}$ by a constant $C_1$, as proposed in previous studies (see for example \cite{Kratz1973,Hornshoj1974}). Possible resonant structures in $S_{\beta}$, considered in e.g. \cite{Klapdor1983, Izosimov2011}, are thus ignored, thereby assuming a limited sensitivity of $T_{\rm 1/2p,\beta DF}$ on $S_{\beta}$ with respect to the scale of the systematic trend discussed here. This approximation is further supported by the study in \cite{Veselsky2012}, which shows a limited influence of $S_{\beta}$ in the calculation of $P_{\beta \mathrm{DF}}$. Furthermore, $C_1$ was taken equal for all isotopes listed in Table \ref{tab:Tbdf}, thereby neglecting possible variations of $C_1$ with respect to the nuclear properties of the $\beta$DF precursors - such as mass, proton number, isospin, spin and parity.\\
The Fermi function $F$ can be fairly well described by the function $C_2(Q_{EC}-E)^2$ \cite{Habs1978,Hall1992,logft} for EC decay. The prefactor $C_2$ was again considered element independent, thereby ignoring its slight dependence on the atomic number $Z$ \cite{logft}.
According to \cite{Firestone1999,logft}, EC decay is dominant for transition energies below 5\,MeV if $Z$ exceeds 80. Since $Q_{\beta}$ values of $\beta$DF precursors in the uranium region are typically smaller than 5\,MeV (see Table \ref{tab:Tbdf}), $\beta^+$ decay can be disregarded here. $Q_{\beta}$ values in the neutron-deficient lead region can however reach 10\,--\,11\,MeV, implying a relatively high $\beta^+$ over EC decay ratio to the ground or a low-lying excited state in the daughter. However, since $\beta$DF should primarily happen at excitation energies which are only a few MeV below $Q_{\beta}$ \cite{Moller2012}, EC-delayed fission should dominate over $\beta^+$ delayed fission in the full region of the nuclear chart (see further).\\
The prompt decay of an excited state in a nucleus can, in the most general case, happen through fission, emission of a $\gamma$ ray, proton, $\alpha$ particle or neutron. The total decay width is thus given by $\Gamma_{\text{tot}} = \Gamma_{\rm f} + \Gamma_{\gamma} + \Gamma_{\rm p} + \Gamma_{\alpha} + \Gamma_{\rm n}$.\\
For the $\beta$DF precursors considered in Table \ref{tab:Tbdf}, the neutron separation energies exceed the $Q_{\beta}$ value by at least several MeV \cite{AME2012} and charge particle-emission is strongly hindered due to the large Coulomb barrier. Therefore, the de-excitation of states below $Q_{\beta}$ is mostly dominated by $\gamma$ decay, which makes that $\Gamma_{\text{tot}} \simeq \Gamma_{\gamma}$ \cite{NRV,Veselsky2012}.
In addition, $\Gamma_{\gamma}$ can be approximated by a constant (see for example \cite{Veselsky2012}). Reference \cite{NRV} provides a calculation of $\Gamma_f$ with respect to the excitation energy $E$ by including the fission-barrier penetrability and the influence of level densities at the ground state and saddle point. This calculation shows that $\Gamma_{\text{f}}$ seems well approximated by a single exponential behavior $\Gamma_{\text{f}} \sim e^{-X(B_{ \rm f}-E)}$ at excitation energies around $B_{\rm f}$. For the fissile nuclei listed in Table \ref{tab:Tbdf}, the decay constant adopts a value $X\approx4$ MeV$^{-1}$ \cite{NRV}. The ratio $\Gamma_{\rm f}/\Gamma_{\rm tot}$ is thus approximated by

\begin{equation}
\frac{\Gamma_{\rm f}}{\Gamma_{\rm tot}}(E) \simeq \frac{\Gamma_{\rm f}}{\Gamma_{\gamma}}(E) \approx C_3e^{-X(B_{ \rm f}-E)}.
\label{ratioG}
\end{equation}

The constants $C_3$ and $X$ are assumed to adopt the same value for all isotopes of interest. At excitation energies $E$ moderately above $B_{\rm f}$, de-excitation by fission should dominate and $\Gamma_{\rm f}$/$\Gamma_{tot}(E)$ will thus be close to unity. Since the $Q_{\beta}$ value of most known $\beta$DF precursors (see Table \ref{tab:Tbdf}) does not exceed $B_{\rm f}$ of the daughter by more than a few MeV, it is further assumed that equation (\ref{ratioG}) remains valid for excitation energies in the daughter nucleus close to $Q_{\beta}$.  \\
Using the above approximations and taking $C=C_1C_2C_3$, the right-hand side of equation (\ref{eq:lambdaBdf}) reduces to

\begin{equation}
\lambda_{\beta \rm DF} = C\int_0^{Q_{\beta}}(Q_{\beta}-E)^2e^{-X(B_{\rm f}-E)}dE.
\label{eq:l_calc}
\end{equation}

\subsection{Calculating $\lambda_{bdf}$}

Equation (\ref{eq:l_calc}) can be rewritten, in order to isolate the exponential dependance on $(Q_{\beta}-B_{\rm f})$, as

\begin{equation}
\lambda_{\beta \rm DF} = Ce^{X(Q_{\beta}-B_{\rm f})}{\int_0^{Q_{\beta}}(Q_{\beta}-E)^2e^{-X(Q_{\beta}-E)}}dE.
\label{eq:l_calc2}
\end{equation}

\begin{figure}
\includegraphics[width=0.5\columnwidth]{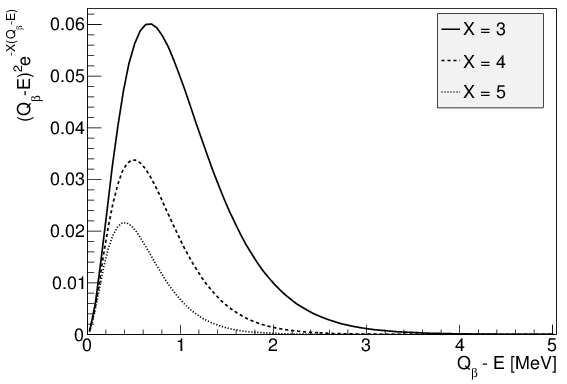}
\caption{Plot showing the integrand of equation (\ref{eq:l_calc2}), which is proportional to the $\beta$DF probability, for $X$ equal to 3,4 or 5.}
\label{fig:prob}
\end{figure}

The integrand in equation (\ref{eq:l_calc2}) is thus proportional to the $\beta$DF probability at a given $E$ of the daughter nucleus. This function, plotted in Figure \ref{fig:prob} for different values of $X$ around the deduced value $X\approx4$\,$\mathrm{MeV}^{-1}$ from \cite{NRV}, shows that $\beta$DF primarily happens at energy levels 0\,--\,2\,MeV below $Q_{\beta}$. Moreover, since all $Q_{\beta}$ values of the neutron-deficient $\beta$DF precursors listed in Table \ref{tab:Tbdf} exceed $\sim$\,2\,MeV, the value of the integral in equation (\ref{eq:l_calc2}) is little dependent on the precise value of $Q_{\beta}$. As a consequence, $\lambda_{\beta \rm DF}$ primarily depends on the difference $(Q_{\beta}-B_{\rm f})$.

In order to prove latter statement analytically, a substitution with $u=X(Q_{\beta}-E)$ and adjustment of integration borders in equation (\ref{eq:l_calc2}) is performed:

\begin{equation}
\lambda_{\beta \rm DF} = \frac{Ce^{X(Q_{\beta}-B_{\rm f})}}{X^3}{\int_0^{XQ_{\beta}}u^2e^{-u}du}.
\label{eq:l_calc3}
\end{equation}

The integral in equation (\ref{eq:l_calc3}) is similar to the mathematical form of the so-called normalized upper incomplete Gamma function, defined as

\begin{equation}
\Gamma(s,x) = \frac{1}{\Gamma(s)}\int^x_0t^{s-1}e^{-t}dt,
\label{GammaFI}
\end{equation}

whereby $\Gamma(s)$ is

\begin{equation}
\Gamma(s) = \int^{+\infty}_0t^{s-1}e^{-t}dt.
\label{GammaFC}
\end{equation}

Equation (\ref{eq:l_calc3}) thus transforms into

\begin{equation}
\lambda_{\beta \rm DF}=\frac{Ce^{X(Q_{\beta}-B_{\rm f})}}{X^3}\Gamma(3)\Gamma(3,XQ_{\beta}).
\label{eq:l_calc4}
\end{equation}

\begin{figure}
\includegraphics[width=0.5\columnwidth]{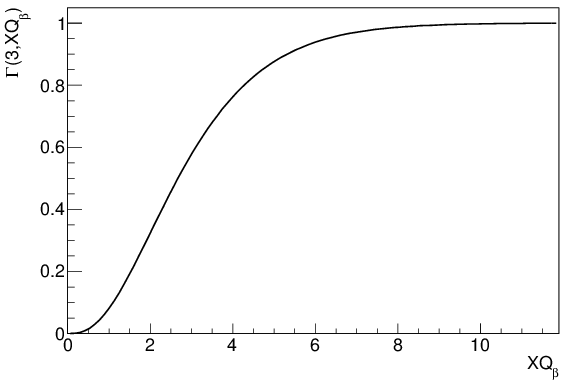}
\caption{The normalized incomplete Gamma function $\Gamma(3,XQ_{\beta})$, needed for the calculation of the integral under the $\beta$DF probability curves shown in Figure \ref{fig:prob}.}
\label{fig:gamma}
\end{figure}

Table \ref{tab:Tbdf} shows that all $Q_\beta$ values of the neutron-deficient $\beta$DF precursors exceed 3\,MeV, while the fitted values for $X$ in Table \ref{tab:FitResults}, as well as the theoretical estimate from \cite{NRV} ($X\approx4$\,$\mathrm{MeV}^{-1}$), are all greater than 1.7\,$\mathrm{MeV}^{-1}$. The value $XQ_{\beta}$ thus exceeds 5 in all discussed cases, implying that, as shown in Figure \ref{fig:gamma}, one can thus safely approximate $\Gamma(3,XQ_{\beta})\simeq 1$ in equation (\ref{eq:l_calc4}).

In this simple picture, it is thus found that $\text{ln}(\lambda_{\beta DF})$ depends linearly on ($Q_{\beta}-B_{\rm f}$). In terms of the partial $\beta$DF half-life $T_{\rm 1/2p,\beta DF}$ one finds the relation

\begin{equation}
\mathrm{log_{10}}(T_{\rm 1/2p,\beta DF}) = C' -  X\mathrm{log_{10}}(e)(Q_{\beta}-B_{\rm f}),
\label{eq:l_def}
\end{equation}

with the constant $C'$ given by

\begin{equation}
C'=\mathrm{ln}\left(\frac{\mathrm{ln(2)}X^3}{C\Gamma(3)}\right)\mathrm{log_{10}}(e).
\end{equation}

\section{Systematic comparison of experimental data}
\label{sec:data}

\begin{figure*}
\includegraphics[width=1.0\columnwidth]{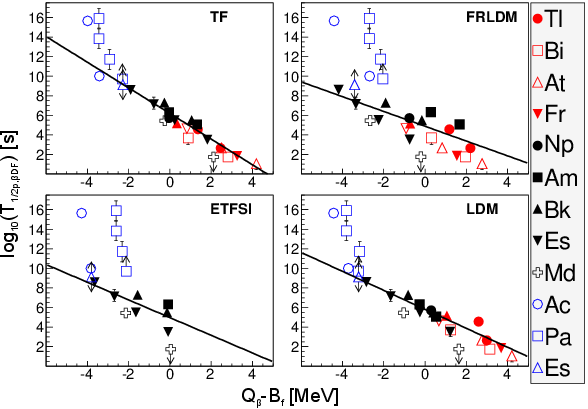}
\caption{(Color online) Plots of $T_{\rm 1/2p,\beta DF}$ versus ($Q_{\beta} - B_{\rm f}$) for different fission models as listed in Table \ref{tab:Tbdf}. The closed symbols, representing reliable values for $T_{1/2p,\beta\mathrm{DF}}$ in Table \ref{tab:Tbdf} are used for a linear fit with equal weights to the data points. Other data from Table \ref{tab:Tbdf} are indicated by the open symbols. The color code represents the different regions of the nuclear chart for which $\beta$DF has been experimentally observed : the neutron-deficient lead region (red), the neutron-deficient (black) and neutron-rich (blue) uranium region.}
\label{fig:Tbdf}
\end{figure*}

This section aims at verifying equation (\ref{eq:l_def}) by using experimental $\beta$DF partial half-lives and theoretical values for ($Q_{\beta}-B_{\rm f}$), summarized in Table \ref{tab:Tbdf} and Figure \ref{fig:Tbdf}. Tabulated fission barriers from four different fission models were used, of which three are based on a macroscopic-microscopic and one a mean-field approach. The latter model is based on the Extended Thomas-Fermi plus Strutinsky Integral (ETFSI) method \cite{Mamdouh2001}, but tabulated barriers for the most neutron-deficient isotopes in Table \ref{tab:Tbdf} are absent in literature. The microscopic-macroscopic approaches all rely on shell corrections from \cite{Moller1995} and describe the macroscopic structure of the nucleus by either a Thomas-Fermi (TF) \cite{Myers1999}, liquid-drop (LDM) \cite{NRV} or the Finite-Range Liquid-Drop Model (FRLDM) \cite{Moller2015}. The $Q_{\beta}$ values were taken from the 2012 atomic mass evaluation tables \cite{AME2012} and are derived from the difference between the atomic masses of parent $M_P(Z,A)$ and daughter $M_D(Z',A)$ nuclei as

\begin{equation}
Q_{\beta} = c^2[M_P(Z,A) - M_D(Z',A)].
\label{eq:Qvalue}
\end{equation}

About half of these values are known from experiments, while the others are deduced from extrapolated atomic masses. In latter cases, the difference of the $Q_{\beta}$ values from \cite{AME2012} with the theoretical values from \cite{Moller1995} or \cite{Myers1996} is always lower than 0.4 MeV.\\

\begingroup
\squeezetable
\begin{table*}
\caption{List of all precursors for which $\beta$DF was observed. The measured half-life $T_{1/2}$, $\beta$-branching ratio $b_{\beta}$, $\beta$DF probability $P_{\beta \mathrm{DF}}$, ratio of observed $\beta$DF to $\alpha$ decays $N_{\beta \mathrm{DF}}/N_{\alpha}$ and calculated $\beta$DF partial half lives $T_{1/2p,\beta\mathrm{DF}}$ are listed. Reliable values for $T_{1/2p,\beta\mathrm{DF}}$, as evaluated by the criteria in \cite{Andreyev2013a}, are indicated in bold. ($Q_{\beta} - B_{\rm f}$) is tabulated for fission barriers from four different fission models : Thomas-Fermi (TF) \cite{Myers1999}, Finite Range Liquid Drop (FRLDM) \cite{Moller2015}, Liquid Drop (LDM) \cite{NRV} and the Extended Thomas-Fermi plus Strutinsky Integral (ETFSI) model \cite{Mamdouh2001}. $Q_{\beta}$ values were taken from \cite{AME2012} and are defined by equation (\ref{eq:Qvalue}).}
\label{tab:Tbdf}
\begin{ruledtabular}
\begin{tabular}{lccccccc D{,}{\times}{-1} D{,}{\times}{-1} D{,}{\times}{-1} c}
&                         &                   & \multicolumn{4}{c}{$Q_{\beta}-B_{\rm f}$ (MeV)} &             &                         &                                    &    &\\
\cline{4-7}
precursor & $T_{1/2}$ (s) & $Q_{\beta}$ (MeV) & TF & FRLDM & LDM & ETFSI                        & $b_{\beta}$ & \multicolumn{1}{c}{$P_{\beta \mathrm{DF}}$} & \multicolumn{1}{c}{$N_{\beta \mathrm{DF}}/N_{\alpha}$} & \multicolumn{1}{c}{$T_{1/2p,\beta\mathrm{DF}}$ (s)}  & ref. \\
\hline
\multicolumn{12}{l}{\it $\beta^+$/EC-delayed fission in the neutron-deficient lead region} \\
$^{178}\mathrm{Tl}$    & 0.25(2)                                                                   &11.5   & 2.5  & 2.2  & 3.0  &      & 0.38(2)
                                                                                       & 1.5(6), 10^{-3}           &                             & \mathbf{4(2)},\mathbf{10^2} & \cite{Liberati2013} \\

$^{180}\mathrm{Tl}$    & 1.09(1)                                                                   &11.0   & 1.4  & 1.2  & 2.6  &      & 0.94(4)
                                                                                       & 3.2(2), 10^{-5}           &                          & \mathbf{3.6(3)},\mathbf{10^4}
& \cite{Elseviers2013} \\
$^{186g,m}\mathrm{Bi}$ & 0.012(3) \footnote{\label{foot:T}Value extracted according to equation (\ref{eq:T1/2}) by using evaluated experimental data from \cite{Nubase2012}.}  &11.6 & 2.8  & 2.0  & 3.1  &      & $\sim 0.006$
\footnote{\label{foot:btheo}Calculated $\beta$-branching ratio from \cite{Moller1997}.} &                                 & 2.2(13), 10^{-4}     & \multicolumn{1}{c}{$56(35)$}    & \cite{Lane2013} \\

$^{188g,m}\mathrm{Bi}$ & 0.16(10) \textsuperscript{\ref{foot:T}}                                   &10.6   & 0.9  & 0.3  & 1.2  &      & $\sim 0.03$
\textsuperscript{\ref{foot:btheo}}                                                     &                                  & 3.2(16),10^{-5}     & 5(4),10^3              & \cite{Lane2013} \\

$^{192g,m}\mathrm{At}$ & 0.05(4) \textsuperscript{\ref{foot:T}}                                    &11.0   & 4.2  & 2.8  & 4.2  &      & $\sim  0.03$
\textsuperscript{\ref{foot:btheo}}                                                     &                                  & 4.2(9), 10^{-3}      & \multicolumn{1}{c}{$12(9)$}                         & \cite{Andreyev2013} \\

$^{194g,m}\mathrm{At}$ & 0.28(3) \textsuperscript{\ref{foot:T}}                                    &10.3   & 2.5  & 0.8  & 2.7  &      & $\sim 0.08$ \textsuperscript{\ref{foot:btheo}}                                                     &                                  & 5.9(4), 10^{-5}      & 4.8(6),10^2 & \cite{Ghys2014} \\

$^{196}\mathrm{At}$    & 0.358(5)                                                                  &9.6    & 0.3  & -0.7 & 1.1  &      & $0.026(1)$
                                                                                       & 9(1), 10^{-5}            & 2.3(2), 10^{-6}      & \mathbf{1.5(2)},\mathbf{10^5} & \cite{Ghys2014,Truesdale2014} \\

$^{200}\mathrm{Fr}$    & 0.049(4) \textsuperscript{\ref{foot:T}}                                   &10.2   & 3.3  & 1.5  & 3.7  &      & $<0.021(4)$
                                                                                       & >3.1(17), 10^{-2}        & 7^{+5}_{-3} ,10^{-4} & \mathbf{7^{+6}_{-3}},\mathbf{10} & \cite{Ghys2014}\\

$^{202g,m}\mathrm{Fr}$ & 0.33(4) \footnote{Value extracted according to equation (\ref{eq:T1/2}) by using experimental data from \cite{Kalaninova2014}.}                &9.4   & 0.8  & -0.9 & 0.7  &      & $\sim 0.007$ \textsuperscript{\ref{foot:btheo}}                                                     &                                & 7.3(8),10^{-7}     & 4.5(8),10^4            & \cite{Ghys2014} \\
\\

\multicolumn{11}{l}{\it $\beta^+$/EC-delayed fission in the neutron-deficient uranium region} \\

$^{228}\mathrm{Np}$    & 61(1)                                                                      &4.4   & 0.0  & -0.8 & 0.3  &      & $0.60(7)$
                                                                                       & 2.0(9), 10^{-4}           &                             & \mathbf{5.1(2)},\mathbf{10^5}     & \cite{Kreek1994} \\

$^{232}\mathrm{Am}$    & 79(2)                                                                      &4.9   & 1.3  & 1.7  & 0.5  &      & $\sim 0.96$ \textsuperscript{\ref{foot:btheo}}                                                     & 6.9(10), 10^{-4}          &                             & \mathbf{1.2(2)},\mathbf{10^{5}}   & \cite{Hall1990} \\

$^{234}\mathrm{Am}$    & 139(5)                                                                     &4.1   & 0.0  & 0.3  & -0.3 & -0.1 & $\sim 1.00$ \textsuperscript{\ref{foot:btheo}}                                                     & 6.6(18), 10^{-5}          &                             & \mathbf{2.1(6)},\mathbf{10^6}     & \cite{Hall1990a} \\

$^{238}\mathrm{Bk}$    & 144(5)                                                                     &4.8   & 1.1  & -0.2 & 0.4  & -0.1 & $\sim 0.95$ \textsuperscript{\ref{foot:btheo}}                                                     & 4.8(20), 10^{-4}          &                             & \mathbf{3.2(13)},\mathbf{10^5}    & \cite{Kreek1994a} \\

$^{240}\mathrm{Bk}$    & 252(48)                                                                    &3.9   & -0.3 & -1.9 & -0.8 & -1.6 & $\sim 1.00$ \textsuperscript{\ref{foot:btheo}}                                                     & 1.3^{+1.8}_{-0.7}, 10^{-5}&                             & \mathbf{1.9^{+2.3}_{-1.1}},\mathbf{10^{7}} & \cite{Galeriu1983} \\

$^{242}\mathrm{Es}$    & 11(3)                                                                      &5.4   & 1.8  & -0.7 & 1.2  & -0.1 & $0.57(3)$
\footnote{$\beta$-branching ratio from \cite{Antalic2009}.}                             & 6(2), 10^{-3}             &                             & \mathbf{3(1)},\mathbf{10^3}         & \cite{Shaughnessy2000} \\

$^{244}\mathrm{Es}$    & 37(4) \textsuperscript{\ref{foot:T}}                                       &4.5   & 0.2  & -2.2 & -0.3 & -1.7 & $0.96(3)$ \footnote{\label{foot:bexp}Evaluated $\beta$-branching ratio from \cite{Nubase2012}.}  & 1.2(4), 10^{-4}           &                             & \mathbf{3(1)},\mathbf{10^5}         & \cite{Shaughnessy2002} \\

$^{246}\mathrm{Es}$    & 462(30)                                                                    &3.8   & -0.8 & -3.4 & -1.7 & -2.7 & $0.901(18)$  \textsuperscript{\ref{foot:bexp}}                                                       & 3.7^{+8.5}_{-3.0}, 10^{-5}&                             & \mathbf{1.4^{+5.9}_{-1.0}},\mathbf{10^{7}}       & \cite{Shaughnessy2001} \\

$^{248}\mathrm{Es}$    & $1.4(2)\times 10^{3}$                                                      &3.1   & -1.9 & -4.2 & -2.8 & -3.6 & $0.997(3)$ \textsuperscript{\ref{foot:bexp}}                                                       & 3.5(18), 10^{-6}          &                             & \mathbf{4.0(21)},\mathbf{10^8}      & \cite{Shaughnessy2001} \\

$^{246m2}\mathrm{Md}$  & \multicolumn{1}{c}{$4.4(8)$}                                               &5.9   & 2.1  & -0.2 & 1.6  & 0.0  & $>0.77$
                                                                                       & \multicolumn{1}{c}{$> 0.1$}            &                             & \multicolumn{1}{c}{$<57$}                & \cite{Antalic2009} \\

$^{250}\mathrm{Md}$    & $52(6)$ \textsuperscript{\ref{foot:T}}                                     &4.6   & -0.3 & -2.7 & -1.0 & -2.1 & $0.93(3)$ \textsuperscript{\ref{foot:bexp}}                                                       & 2^{+2}_{-1}, 10^{-4}       &                             & 3^{+3}_{-1},10^5 & \cite{Gangrsky1980} \\
\\

\multicolumn{11}{l}{\it $\beta^-$-delayed fission in the neutron-rich uranium region} \\

$^{228}\mathrm{Ac}$    & $2.214(7)\times 10^4$ \textsuperscript{\ref{foot:T}}                       &2.1   & -4.0 & -4.4 & -4.4 & -4.3 & $\sim 1.00$ \textsuperscript{\ref{foot:btheo}}                                                    & 5(2), 10^{-12}             &                             & 4(2),10^{15} & \cite{Yanbing2006}\\

$^{230}\mathrm{Ac}$    & $122(3)$ \textsuperscript{\ref{foot:T}}                                     &3.0  & -3.4 & -2.7 & -3.7 & -3.8 & $\sim 1.00$ \textsuperscript{\ref{foot:btheo}}                                                    & 1.19(40), 10^{-9}          &                             & 1.0(3), 10^{10} & \cite{Shuanggui2001} \\

$^{234g}\mathrm{Pa}$   & $2.41(2)\times 10^4$ \textsuperscript{\ref{foot:T}}                         &2.2  & -3.4 & -2.7 & -3.8 & -2.6 & $\sim 1.00$ \textsuperscript{\ref{foot:btheo}}                                                    & 3, 10^{-(12\pm1)}          &                             & 8, 10^{(15\pm 1)} & \cite{Gangrsky1978} \\

$^{234m}\mathrm{Pa}$   & $69.54(66)$ \textsuperscript{\ref{foot:T}}                                  &2.2  & -3.4 & -2.7 & -3.8 & -2.6 & $0.9984(4)$
                                                                                      & \multicolumn{1}{c}{$10^{-(12\pm 1)}$} &                             & 7,10^{(13\pm 1)} & \cite{Gangrsky1978} \\

$^{236}\mathrm{Pa}$    & $546(6)$ \textsuperscript{\ref{foot:T}}                                     &2.9  & -2.9 & -2.1 & -3.2 & -2.3 & $\sim 1.00$ \textsuperscript{\ref{foot:btheo}}                                                    & \multicolumn{1}{c}{$10^{-9\pm 1}$}                    &                             & 5, 10^{(11\pm 1)} & \cite{Gangrsky1978} \\

$^{238}\mathrm{Pa}$    & $138(6)$ \textsuperscript{\ref{foot:T}}                                     &3.6  & -2.3 & -2.0 & -3.2 & -2.1 & $\sim 1.00$ \textsuperscript{\ref{foot:btheo}}                                                    & <2.6,10^{-8}            &                             & >5.3,10^9   & \cite{Baas-May1985} \\

$^{256m}\mathrm{Es}$   & $2.7\times 10^4$ \textsuperscript{\ref{foot:T}}                             &1.7  & -2.3 & -3.4 & -3.2 & -3.8 & $\sim 1.00$ \textsuperscript{\ref{foot:btheo}}                                                    & \sim 2,10^{-5}            &                             & \sim 1,10^9  & \cite{Hall1989} \\

\end{tabular}
\end{ruledtabular}
\end{table*}
\endgroup

$T_{\rm 1/2p,\beta DF}$ values were extracted from reported $P_{\beta \mathrm{DF}}$ values using equation (\ref{eq:PbdfvsTbdf}), if the precursor nucleus has a significant $\beta$-decay branch ($b_{\beta} \gtrsim 10$\,\%). When multiple measurements on $P_{\beta \mathrm{DF}}$ were performed, only the reliable value, as evaluated by \cite{Andreyev2013a}, or the most recent value was tabulated. In case of a dominant $\alpha$-decay branch ($b_{\beta} \lesssim 10$\,\%), $T_{\rm 1/2p,\beta DF}$ was calculated by equation (\ref{eq:Tbdf}), whereby $N_{\mathrm{dec,tot}}$ was approximated by the observed amount of $\alpha$ decays $N_{\alpha}$, corrected for detection efficiency.\\
Since the isotopes $^{186,188}\mathrm{Bi}$, $^{192,194}\mathrm{At}$ and $^{202}\mathrm{Fr}$ have both a ground and a low-lying alpha-decaying isomeric state with comparable half-lives, only an overall $N_{\beta \mathrm{DF}}/N_{\alpha}$ value could be extracted with present experimental techniques. We refer the reader for a detailed discussion of this issue to \cite{Lane2013,Andreyev2013,Ghys2014}. Therefore, these precursors have been excluded from the fit in Figure \ref{fig:Tbdf}. Nonetheless, as a first approximation the value for $T_{1/2p,\beta\mathrm{DF}}$ was extracted by defining the half-life $T_{1/2}$, shown in table \ref{tab:Tbdf}, as the unweighted average

\begin{equation}
T_{1/2} = \frac{T_{\rm 1/2,g} + T_{\rm 1/2,m}}{2}.
\label{eq:T1/2}
\end{equation}

where the respective half-lives for ground and isomeric states are denoted by $T_{\rm 1/2,g}$ and $T_{\rm 1/2,m}$. The uncertainty $\Delta T_{1/2}$ is conservatively taken as

\begin{equation}
\Delta T_{1/2} = \frac{|T_{\rm 1/2,g} - T_{\rm 1/2,m}|}{2}.
\label{eq:DT1/2}
\end{equation}

Figure \ref{fig:Tbdf} shows $\mathrm{log_{10}}(T_{\rm 1/2p,\beta DF})$ against ($Q_{\beta}-B_{\rm f}$) for the fission barriers from the four different models under consideration. Using the same evaluation criteria as proposed in \cite{Andreyev2013a} for $P_{\beta \mathrm{DF}}$ measurements, 13 reliable $T_{1/2p,\beta\mathrm{DF}}$ values, marked in bold in Table \ref{tab:Tbdf}, were selected. These data points, represented by the closed symbols, are fitted by a linear function. An equal weight to all fit points is given because the experimental uncertainties on $\mathrm{log_{10}}(T_{\rm 1/2p,\beta DF})$ are in most cases much smaller than the deviation of the data points with the fitted line, of which the extracted parameters are summarized in Table \ref{tab:FitResults}. The remaining data points from Table \ref{tab:Tbdf} are shown by open symbols and were excluded from the fit. The color code discriminates between the neutron-deficient lead region (red), neutron-deficient (black) and neutron-rich (blue) uranium region. \\

Figure \ref{fig:Tbdf} illustrates a linear dependence of $\mathrm{log_{10}}(T_{\rm 1/2p,\beta DF})$ on $(Q_{\beta}-B_{\rm f})$ for TF and LDM barriers for over 7 orders of magnitude of $T_{1/2p,\beta\mathrm{DF}}$. In addition, Table \ref{tab:FitResults} shows a relatively small root-mean-square deviation (RMSD) of the 13 reliable experimental $\mathrm{log_{10}}(T_{\rm 1/2p,\beta DF})$ values (represented by the closed symbols in Figure \ref{fig:Tbdf}) to the corresponding values extracted from the fit. The dependence is somewhat less pronounced for the FRLDM model, as evidenced by a larger RMSD value. A similar linear trend is observed for the ETFSI model, but the lack of tabulated fission barriers in the neutron-deficient region, especially in the lead region, prohibits drawing definite conclusions.\\
Moreover, Table \ref{tab:FitResults} shows that the four fitted values for $X$ are similar to each other as well as to the theoretical estimate $X \approx 4$\,$\mathrm{MeV}^{-1}$ \cite{NRV}. The extracted values for the offset $C'$ are also found to be comparable.\\

\begin{table}
\caption{Results of the fits, corresponding to four different fission models, shown in Figure \ref{fig:Tbdf}. The values for the parameters $X$ and $C'$ in equation (\ref{eq:l_def}) are listed. Also the root-mean-square deviations (RMSD) of the reliable experimental $\mathrm{log_{10}}(T_{\rm 1/2p,\beta DF})$ values (represented by the closed symbols in Figure \ref{fig:Tbdf}) to the fit are given.}
\label{tab:FitResults}
\begin{ruledtabular}
\begin{tabular}{cccc}
Model & X ($\mathrm{MeV}^{-1}$) & C' (MeV) & RMSD\\
\hline
TF    & 3.0(2)                  & 6.2(1)   & 0.47\\
FRLDM & 1.7(4)                  & 4.9(3)   & 1.19\\
ETFSI & 2.1(7)                  & 5.0(6)   & 1.10\\
LDM   & 2.2(2)                  & 5.8(2)   & 0.62\\
\end{tabular}
\end{ruledtabular}
\end{table}

In contrast to a rather good agreement for most neutron-deficient nuclei, all models show a larger systematical deviation from this linear trend for the neutron-rich $\beta$DF precursors $^{228}\mathrm{Ac}$ and $^{234,236}\mathrm{Pa}$. In \cite{Andreyev2013a}, concerns were raised on the accuracy of the $P_{\beta \mathrm{DF}}$ values measured in this region, which could explain this deviation. Note also that the precursors in this region of the nuclear chart undergo $\beta^{-}$ decay in contrast to the EC-delayed fission on the neutron-deficient side for which equation (\ref{eq:l_def}) was deduced, influencing the numeric value of the offset $C'$. In particular, the Fermi function for $\beta^{-}$ decay is approximately proportional to $(Q_{\beta}-E)^5$ \cite{Hall1992,Kuznetsov1999}, in contrast to the quadratic dependence on $(Q_{\beta}-E)$ for EC decay. The parameter $X$ should however remain unchanged, because equation (\ref{ratioG}) approximating $\Gamma_{\rm f}/\Gamma_{\rm tot}$ remains valid as long as the neutron-separation energy $S_{\rm n}$ is larger than $Q_{\beta}$. Since, at excitation energies higher than $S_{\rm n}$, de-excitation through neutron emission is favored over decay by $\gamma$-ray emission, thus implying $\Gamma_{\text{tot}} \simeq \Gamma_{\rm n} \gg \Gamma_{\gamma},\Gamma_{\rm f}$ \cite{Bohr1939,NRV}. For all nuclei mentioned in Table \ref{tab:Tbdf} however, $Q_{\beta}$ is below $S_{\rm n}$. An approximation of $T_{1/2p,\beta\mathrm{DF}}$, similar to equation (\ref{eq:l_def}), can thus also be derived for neutron-rich $\beta$DF precursors by taking into account above considerations. However, considering the limited experimental information on $\beta$DF in the neutron-rich region, a detailed derivation is omitted in this paper.

\section{Conclusions}

Recent experiments have measured the $\beta$DF of 9 precursor nuclei in the neutron-deficient lead region. Because of the dominant $\alpha$-decay branch in most of these nuclei, $\beta$DF probabilities  are extracted with large experimental uncertainties. In contrast, the partial half-life for $\beta$DF can be determined with a better accuracy. In addition, $T_{\rm 1/2p,\beta DF}$ can be easily derived from the $\beta$DF probability by using equation (\ref{eq:PbdfvsTbdf}).\\
A systematical evaluation of $\beta$DF partial half-lives was performed by using fission barriers deduced from four different models for a broad range of nuclei in the lead and uranium regions. A linear relation between $\mathrm{log_{10}}(T_{\rm 1/2p,\beta DF})$ and ($Q_{\beta}-B_{\rm f}$) was observed for neutron-deficient precursor nuclei, when using tabulated fission barriers from the TF or LDM approach, and to a lesser extent for FRLDM and ETFSI barriers. This linear trend persists for values of $T_{\rm 1/2p,\beta DF}$ spanning over 7 orders of magnitude and a wide variety of precursor nuclei going from $^{178}\mathrm{Tl}$ to $^{248}\mathrm{Es}$ with $N$/$Z$ ratios of 1.20 and 1.51 respectively. This observation may help to assess $\beta$DF branching-ratios in very neutron-rich nuclei, which are inaccessible by present experimental techniques but might play a role in the fission-recycling mechanism of the r-process nucleosynthesis.\\
\\
This work has been funded by FWO-Vlaanderen (Belgium), by the Slovak Research and Development Agency (Contract No. APVV-0105-10), by the UK Science and Technology Facilities Council (STFC), by the Slovak grant agency VEGA (contract No. 1/0576/13), by the Reimei Foundation of JAEA, and by the European Commission within the Seventh Framework Programme through I3-ENSAR (Contract No. RII3-CT-2010-262010).

\FloatBarrier
\bibliography{Bibliography}

\begin{thebibliography}{48}%
\makeatletter
\providecommand \@ifxundefined [1]{%
 \@ifx{#1\undefined}
}%
\providecommand \@ifnum [1]{%
 \ifnum #1\expandafter \@firstoftwo
 \else \expandafter \@secondoftwo
 \fi
}%
\providecommand \@ifx [1]{%
 \ifx #1\expandafter \@firstoftwo
 \else \expandafter \@secondoftwo
 \fi
}%
\providecommand \natexlab [1]{#1}%
\providecommand \enquote  [1]{``#1''}%
\providecommand \bibnamefont  [1]{#1}%
\providecommand \bibfnamefont [1]{#1}%
\providecommand \citenamefont [1]{#1}%
\providecommand \href@noop [0]{\@secondoftwo}%
\providecommand \href [0]{\begingroup \@sanitize@url \@href}%
\providecommand \@href[1]{\@@startlink{#1}\@@href}%
\providecommand \@@href[1]{\endgroup#1\@@endlink}%
\providecommand \@sanitize@url [0]{\catcode `\\12\catcode `\$12\catcode
  `\&12\catcode `\#12\catcode `\^12\catcode `\_12\catcode `\%12\relax}%
\providecommand \@@startlink[1]{}%
\providecommand \@@endlink[0]{}%
\providecommand \url  [0]{\begingroup\@sanitize@url \@url }%
\providecommand \@url [1]{\endgroup\@href {#1}{\urlprefix }}%
\providecommand \urlprefix  [0]{URL }%
\providecommand \Eprint [0]{\href }%
\providecommand \doibase [0]{http://dx.doi.org/}%
\providecommand \selectlanguage [0]{\@gobble}%
\providecommand \bibinfo  [0]{\@secondoftwo}%
\providecommand \bibfield  [0]{\@secondoftwo}%
\providecommand \translation [1]{[#1]}%
\providecommand \BibitemOpen [0]{}%
\providecommand \bibitemStop [0]{}%
\providecommand \bibitemNoStop [0]{.\EOS\space}%
\providecommand \EOS [0]{\spacefactor3000\relax}%
\providecommand \BibitemShut  [1]{\csname bibitem#1\endcsname}%
\let\auto@bib@innerbib\@empty
\bibitem [{\citenamefont {Andreyev}\ \emph {et~al.}(2010)\citenamefont
  {Andreyev}, \citenamefont {Elseviers}, \citenamefont {Huyse}, \citenamefont
  {{Van Duppen}}, \citenamefont {Antalic}, \citenamefont {Barzakh},
  \citenamefont {Bree}, \citenamefont {Cocolios}, \citenamefont {Comas},
  \citenamefont {Diriken}, \citenamefont {Fedorov}, \citenamefont {Fedosseev},
  \citenamefont {Franchoo}, \citenamefont {Heredia}, \citenamefont {Ivanov},
  \citenamefont {K\"{o}ster}, \citenamefont {Marsh}, \citenamefont {Nishio},
  \citenamefont {Page}, \citenamefont {Patronis}, \citenamefont {Seliverstov},
  \citenamefont {Tsekhanovich}, \citenamefont {{Van den Bergh}}, \citenamefont
  {{Van De Walle}}, \citenamefont {Venhart}, \citenamefont {Vermote},
  \citenamefont {Veselsky}, \citenamefont {Wagemans}, \citenamefont {Ichikawa},
  \citenamefont {Iwamoto}, \citenamefont {M\"{o}ller},\ and\ \citenamefont
  {Sierk}}]{Andreyev2010}%
  \BibitemOpen
  \bibfield  {author} {\bibinfo {author} {\bibfnamefont {A.~N.}\ \bibnamefont
  {Andreyev}}, \bibinfo {author} {\bibfnamefont {J.}~\bibnamefont {Elseviers}},
  \bibinfo {author} {\bibfnamefont {M.}~\bibnamefont {Huyse}}, \bibinfo
  {author} {\bibfnamefont {P.}~\bibnamefont {{Van Duppen}}}, \bibinfo {author}
  {\bibfnamefont {S.}~\bibnamefont {Antalic}}, \bibinfo {author} {\bibfnamefont
  {A.}~\bibnamefont {Barzakh}}, \bibinfo {author} {\bibfnamefont
  {N.}~\bibnamefont {Bree}}, \bibinfo {author} {\bibfnamefont {T.~E.}\
  \bibnamefont {Cocolios}}, \bibinfo {author} {\bibfnamefont {V.~F.}\
  \bibnamefont {Comas}}, \bibinfo {author} {\bibfnamefont {J.}~\bibnamefont
  {Diriken}}, \bibinfo {author} {\bibfnamefont {D.}~\bibnamefont {Fedorov}},
  \bibinfo {author} {\bibfnamefont {V.~N.}\ \bibnamefont {Fedosseev}}, \bibinfo
  {author} {\bibfnamefont {S.}~\bibnamefont {Franchoo}}, \bibinfo {author}
  {\bibfnamefont {J.~A.}\ \bibnamefont {Heredia}}, \bibinfo {author}
  {\bibfnamefont {O.}~\bibnamefont {Ivanov}}, \bibinfo {author} {\bibfnamefont
  {U.}~\bibnamefont {K\"{o}ster}}, \bibinfo {author} {\bibfnamefont {B.~A.}\
  \bibnamefont {Marsh}}, \bibinfo {author} {\bibfnamefont {K.}~\bibnamefont
  {Nishio}}, \bibinfo {author} {\bibfnamefont {R.~D.}\ \bibnamefont {Page}},
  \bibinfo {author} {\bibfnamefont {N.}~\bibnamefont {Patronis}}, \bibinfo
  {author} {\bibfnamefont {M.}~\bibnamefont {Seliverstov}}, \bibinfo {author}
  {\bibfnamefont {I.}~\bibnamefont {Tsekhanovich}}, \bibinfo {author}
  {\bibfnamefont {P.}~\bibnamefont {{Van den Bergh}}}, \bibinfo {author}
  {\bibfnamefont {J.}~\bibnamefont {{Van De Walle}}}, \bibinfo {author}
  {\bibfnamefont {M.}~\bibnamefont {Venhart}}, \bibinfo {author} {\bibfnamefont
  {S.}~\bibnamefont {Vermote}}, \bibinfo {author} {\bibfnamefont
  {M.}~\bibnamefont {Veselsky}}, \bibinfo {author} {\bibfnamefont
  {C.}~\bibnamefont {Wagemans}}, \bibinfo {author} {\bibfnamefont
  {T.}~\bibnamefont {Ichikawa}}, \bibinfo {author} {\bibfnamefont
  {A.}~\bibnamefont {Iwamoto}}, \bibinfo {author} {\bibfnamefont
  {P.}~\bibnamefont {M\"{o}ller}}, \ and\ \bibinfo {author} {\bibfnamefont
  {A.~J.}\ \bibnamefont {Sierk}},\ }\href {\doibase
  10.1103/PhysRevLett.105.252502} {\bibfield  {journal} {\bibinfo  {journal}
  {Phys. Rev. Lett.}\ }\textbf {\bibinfo {volume} {105}},\ \bibinfo {pages}
  {252502} (\bibinfo {year} {2010})}\BibitemShut {NoStop}%
\bibitem [{\citenamefont {Elseviers}\ \emph {et~al.}(2013)\citenamefont
  {Elseviers}, \citenamefont {Andreyev}, \citenamefont {Huyse}, \citenamefont
  {{Van Duppen}}, \citenamefont {Antalic}, \citenamefont {Barzakh},
  \citenamefont {Bree}, \citenamefont {Cocolios}, \citenamefont {Comas},
  \citenamefont {Diriken}, \citenamefont {Fedorov}, \citenamefont {Fedosseev},
  \citenamefont {Franchoo}, \citenamefont {Ghys}, \citenamefont {Heredia},
  \citenamefont {Ivanov}, \citenamefont {K\"{o}ster}, \citenamefont {Marsh},
  \citenamefont {Nishio}, \citenamefont {Page}, \citenamefont {Patronis},
  \citenamefont {Seliverstov}, \citenamefont {Tsekhanovich}, \citenamefont
  {{Van den Bergh}}, \citenamefont {{Van De Walle}}, \citenamefont {Venhart},
  \citenamefont {Vermote}, \citenamefont {Veselsk\'{y}},\ and\ \citenamefont
  {Wagemans}}]{Elseviers2013}%
  \BibitemOpen
  \bibfield  {author} {\bibinfo {author} {\bibfnamefont {J.}~\bibnamefont
  {Elseviers}}, \bibinfo {author} {\bibfnamefont {A.~N.}\ \bibnamefont
  {Andreyev}}, \bibinfo {author} {\bibfnamefont {M.}~\bibnamefont {Huyse}},
  \bibinfo {author} {\bibfnamefont {P.}~\bibnamefont {{Van Duppen}}}, \bibinfo
  {author} {\bibfnamefont {S.}~\bibnamefont {Antalic}}, \bibinfo {author}
  {\bibfnamefont {A.}~\bibnamefont {Barzakh}}, \bibinfo {author} {\bibfnamefont
  {N.}~\bibnamefont {Bree}}, \bibinfo {author} {\bibfnamefont {T.~E.}\
  \bibnamefont {Cocolios}}, \bibinfo {author} {\bibfnamefont {V.~F.}\
  \bibnamefont {Comas}}, \bibinfo {author} {\bibfnamefont {J.}~\bibnamefont
  {Diriken}}, \bibinfo {author} {\bibfnamefont {D.}~\bibnamefont {Fedorov}},
  \bibinfo {author} {\bibfnamefont {V.~N.}\ \bibnamefont {Fedosseev}}, \bibinfo
  {author} {\bibfnamefont {S.}~\bibnamefont {Franchoo}}, \bibinfo {author}
  {\bibfnamefont {L.}~\bibnamefont {Ghys}}, \bibinfo {author} {\bibfnamefont
  {J.~A.}\ \bibnamefont {Heredia}}, \bibinfo {author} {\bibfnamefont
  {O.}~\bibnamefont {Ivanov}}, \bibinfo {author} {\bibfnamefont
  {U.}~\bibnamefont {K\"{o}ster}}, \bibinfo {author} {\bibfnamefont {B.~A.}\
  \bibnamefont {Marsh}}, \bibinfo {author} {\bibfnamefont {K.}~\bibnamefont
  {Nishio}}, \bibinfo {author} {\bibfnamefont {R.~D.}\ \bibnamefont {Page}},
  \bibinfo {author} {\bibfnamefont {N.}~\bibnamefont {Patronis}}, \bibinfo
  {author} {\bibfnamefont {M.~D.}\ \bibnamefont {Seliverstov}}, \bibinfo
  {author} {\bibfnamefont {I.}~\bibnamefont {Tsekhanovich}}, \bibinfo {author}
  {\bibfnamefont {P.}~\bibnamefont {{Van den Bergh}}}, \bibinfo {author}
  {\bibfnamefont {J.}~\bibnamefont {{Van De Walle}}}, \bibinfo {author}
  {\bibfnamefont {M.}~\bibnamefont {Venhart}}, \bibinfo {author} {\bibfnamefont
  {S.}~\bibnamefont {Vermote}}, \bibinfo {author} {\bibfnamefont
  {M.}~\bibnamefont {Veselsk\'{y}}}, \ and\ \bibinfo {author} {\bibfnamefont
  {C.}~\bibnamefont {Wagemans}},\ }\href
  {http://link.aps.org/doi/10.1103/PhysRevC.88.044321} {\bibfield  {journal}
  {\bibinfo  {journal} {Phys. Rev. C}\ }\textbf {\bibinfo {volume} {88}},\
  \bibinfo {pages} {044321} (\bibinfo {year} {2013})}\BibitemShut {NoStop}%
\bibitem [{\citenamefont {Liberati}\ \emph {et~al.}(2013)\citenamefont
  {Liberati}, \citenamefont {Andreyev}, \citenamefont {Antalic}, \citenamefont
  {Barzakh}, \citenamefont {Cocolios}, \citenamefont {Elseviers}, \citenamefont
  {Fedorov}, \citenamefont {Fedoseeev}, \citenamefont {Huyse}, \citenamefont
  {Joss}, \citenamefont {Kalaninov\'{a}}, \citenamefont {K\"{o}ster},
  \citenamefont {Lane}, \citenamefont {Marsh}, \citenamefont {Mengoni},
  \citenamefont {Molkanov}, \citenamefont {Nishio}, \citenamefont {Page},
  \citenamefont {Patronis}, \citenamefont {Pauwels}, \citenamefont {Radulov},
  \citenamefont {Seliverstov}, \citenamefont {Sj\"{o}din}, \citenamefont
  {Tsekhanovich}, \citenamefont {{Van den Bergh}}, \citenamefont {{Van
  Duppen}}, \citenamefont {Venhart},\ and\ \citenamefont
  {Veselsk\'{y}}}]{Liberati2013}%
  \BibitemOpen
  \bibfield  {author} {\bibinfo {author} {\bibfnamefont {V.}~\bibnamefont
  {Liberati}}, \bibinfo {author} {\bibfnamefont {A.~N.}\ \bibnamefont
  {Andreyev}}, \bibinfo {author} {\bibfnamefont {S.}~\bibnamefont {Antalic}},
  \bibinfo {author} {\bibfnamefont {A.}~\bibnamefont {Barzakh}}, \bibinfo
  {author} {\bibfnamefont {T.~E.}\ \bibnamefont {Cocolios}}, \bibinfo {author}
  {\bibfnamefont {J.}~\bibnamefont {Elseviers}}, \bibinfo {author}
  {\bibfnamefont {D.}~\bibnamefont {Fedorov}}, \bibinfo {author} {\bibfnamefont
  {V.~N.}\ \bibnamefont {Fedoseeev}}, \bibinfo {author} {\bibfnamefont
  {M.}~\bibnamefont {Huyse}}, \bibinfo {author} {\bibfnamefont {D.~T.}\
  \bibnamefont {Joss}}, \bibinfo {author} {\bibfnamefont {Z.}~\bibnamefont
  {Kalaninov\'{a}}}, \bibinfo {author} {\bibfnamefont {U.}~\bibnamefont
  {K\"{o}ster}}, \bibinfo {author} {\bibfnamefont {J.~F.~W.}\ \bibnamefont
  {Lane}}, \bibinfo {author} {\bibfnamefont {B.}~\bibnamefont {Marsh}},
  \bibinfo {author} {\bibfnamefont {D.}~\bibnamefont {Mengoni}}, \bibinfo
  {author} {\bibfnamefont {P.}~\bibnamefont {Molkanov}}, \bibinfo {author}
  {\bibfnamefont {K.}~\bibnamefont {Nishio}}, \bibinfo {author} {\bibfnamefont
  {R.~D.}\ \bibnamefont {Page}}, \bibinfo {author} {\bibfnamefont
  {N.}~\bibnamefont {Patronis}}, \bibinfo {author} {\bibfnamefont
  {D.}~\bibnamefont {Pauwels}}, \bibinfo {author} {\bibfnamefont
  {D.}~\bibnamefont {Radulov}}, \bibinfo {author} {\bibfnamefont
  {M.}~\bibnamefont {Seliverstov}}, \bibinfo {author} {\bibfnamefont
  {M.}~\bibnamefont {Sj\"{o}din}}, \bibinfo {author} {\bibfnamefont
  {I.}~\bibnamefont {Tsekhanovich}}, \bibinfo {author} {\bibfnamefont
  {P.}~\bibnamefont {{Van den Bergh}}}, \bibinfo {author} {\bibfnamefont
  {P.}~\bibnamefont {{Van Duppen}}}, \bibinfo {author} {\bibfnamefont
  {M.}~\bibnamefont {Venhart}}, \ and\ \bibinfo {author} {\bibfnamefont
  {M.}~\bibnamefont {Veselsk\'{y}}},\ }\href {\doibase
  10.1103/PhysRevC.88.044322} {\bibfield  {journal} {\bibinfo  {journal} {Phys.
  Rev. C}\ }\textbf {\bibinfo {volume} {88}},\ \bibinfo {pages} {044322}
  (\bibinfo {year} {2013})}\BibitemShut {NoStop}%
\bibitem [{\citenamefont {Ghys}\ \emph {et~al.}(2014)\citenamefont {Ghys},
  \citenamefont {Andreyev}, \citenamefont {Huyse}, \citenamefont {{Van
  Duppen}}, \citenamefont {Sels}, \citenamefont {Andel}, \citenamefont
  {Antalic}, \citenamefont {Barzakh}, \citenamefont {Capponi}, \citenamefont
  {Cocolios}, \citenamefont {Derkx}, \citenamefont {{De Witte}}, \citenamefont
  {Elseviers}, \citenamefont {Fedorov}, \citenamefont {Fedosseev},
  \citenamefont {Hessberger}, \citenamefont {Kalaninov\'{a}}, \citenamefont
  {K\"{o}ster}, \citenamefont {Lane}, \citenamefont {Liberati}, \citenamefont
  {Lynch}, \citenamefont {Marsh}, \citenamefont {Mitsuoka}, \citenamefont
  {M\"{o}ller}, \citenamefont {Nagame}, \citenamefont {Nishio}, \citenamefont
  {Ota}, \citenamefont {Pauwels}, \citenamefont {Page}, \citenamefont
  {Popescu}, \citenamefont {Radulov}, \citenamefont {Rajabali}, \citenamefont
  {Randrup}, \citenamefont {Rapisarda}, \citenamefont {Rothe}, \citenamefont
  {Sandhu}, \citenamefont {Seliverstov}, \citenamefont {Sj\"{o}din},
  \citenamefont {Truesdale}, \citenamefont {{Van Beveren}}, \citenamefont {{Van
  den Bergh}}, \citenamefont {Wakabayashi},\ and\ \citenamefont
  {Warda}}]{Ghys2014}%
  \BibitemOpen
  \bibfield  {author} {\bibinfo {author} {\bibfnamefont {L.}~\bibnamefont
  {Ghys}}, \bibinfo {author} {\bibfnamefont {A.~N.}\ \bibnamefont {Andreyev}},
  \bibinfo {author} {\bibfnamefont {M.}~\bibnamefont {Huyse}}, \bibinfo
  {author} {\bibfnamefont {P.}~\bibnamefont {{Van Duppen}}}, \bibinfo {author}
  {\bibfnamefont {S.}~\bibnamefont {Sels}}, \bibinfo {author} {\bibfnamefont
  {B.}~\bibnamefont {Andel}}, \bibinfo {author} {\bibfnamefont
  {S.}~\bibnamefont {Antalic}}, \bibinfo {author} {\bibfnamefont
  {A.}~\bibnamefont {Barzakh}}, \bibinfo {author} {\bibfnamefont
  {L.}~\bibnamefont {Capponi}}, \bibinfo {author} {\bibfnamefont {T.~E.}\
  \bibnamefont {Cocolios}}, \bibinfo {author} {\bibfnamefont {X.}~\bibnamefont
  {Derkx}}, \bibinfo {author} {\bibfnamefont {H.}~\bibnamefont {{De Witte}}},
  \bibinfo {author} {\bibfnamefont {J.}~\bibnamefont {Elseviers}}, \bibinfo
  {author} {\bibfnamefont {D.~V.}\ \bibnamefont {Fedorov}}, \bibinfo {author}
  {\bibfnamefont {V.~N.}\ \bibnamefont {Fedosseev}}, \bibinfo {author}
  {\bibfnamefont {F.~P.}\ \bibnamefont {Hessberger}}, \bibinfo {author}
  {\bibfnamefont {Z.}~\bibnamefont {Kalaninov\'{a}}}, \bibinfo {author}
  {\bibfnamefont {U.}~\bibnamefont {K\"{o}ster}}, \bibinfo {author}
  {\bibfnamefont {J.~F.~W.}\ \bibnamefont {Lane}}, \bibinfo {author}
  {\bibfnamefont {V.}~\bibnamefont {Liberati}}, \bibinfo {author}
  {\bibfnamefont {K.~M.}\ \bibnamefont {Lynch}}, \bibinfo {author}
  {\bibfnamefont {B.~A.}\ \bibnamefont {Marsh}}, \bibinfo {author}
  {\bibfnamefont {S.}~\bibnamefont {Mitsuoka}}, \bibinfo {author}
  {\bibfnamefont {P.}~\bibnamefont {M\"{o}ller}}, \bibinfo {author}
  {\bibfnamefont {Y.}~\bibnamefont {Nagame}}, \bibinfo {author} {\bibfnamefont
  {K.}~\bibnamefont {Nishio}}, \bibinfo {author} {\bibfnamefont
  {S.}~\bibnamefont {Ota}}, \bibinfo {author} {\bibfnamefont {D.}~\bibnamefont
  {Pauwels}}, \bibinfo {author} {\bibfnamefont {R.~D.}\ \bibnamefont {Page}},
  \bibinfo {author} {\bibfnamefont {L.}~\bibnamefont {Popescu}}, \bibinfo
  {author} {\bibfnamefont {D.}~\bibnamefont {Radulov}}, \bibinfo {author}
  {\bibfnamefont {M.~M.}\ \bibnamefont {Rajabali}}, \bibinfo {author}
  {\bibfnamefont {J.}~\bibnamefont {Randrup}}, \bibinfo {author} {\bibfnamefont
  {E.}~\bibnamefont {Rapisarda}}, \bibinfo {author} {\bibfnamefont
  {S.}~\bibnamefont {Rothe}}, \bibinfo {author} {\bibfnamefont
  {K.}~\bibnamefont {Sandhu}}, \bibinfo {author} {\bibfnamefont {M.~D.}\
  \bibnamefont {Seliverstov}}, \bibinfo {author} {\bibfnamefont {A.~M.}\
  \bibnamefont {Sj\"{o}din}}, \bibinfo {author} {\bibfnamefont {V.~L.}\
  \bibnamefont {Truesdale}}, \bibinfo {author} {\bibfnamefont {C.}~\bibnamefont
  {{Van Beveren}}}, \bibinfo {author} {\bibfnamefont {P.}~\bibnamefont {{Van
  den Bergh}}}, \bibinfo {author} {\bibfnamefont {Y.}~\bibnamefont
  {Wakabayashi}}, \ and\ \bibinfo {author} {\bibfnamefont {M.}~\bibnamefont
  {Warda}},\ }\href {\doibase 10.1103/PhysRevC.90.041301} {\bibfield  {journal}
  {\bibinfo  {journal} {Phys. Rev. C}\ }\textbf {\bibinfo {volume} {90}},\
  \bibinfo {pages} {041301} (\bibinfo {year} {2014})}\BibitemShut {NoStop}%
\bibitem [{\citenamefont {Andreyev}\ \emph
  {et~al.}(2013{\natexlab{a}})\citenamefont {Andreyev}, \citenamefont
  {Antalic}, \citenamefont {Ackermann}, \citenamefont {Bianco}, \citenamefont
  {Franchoo}, \citenamefont {Heinz}, \citenamefont {Hessberger}, \citenamefont
  {Hofmann}, \citenamefont {Huyse}, \citenamefont {Kalaninov\'{a}},
  \citenamefont {Kojouharov}, \citenamefont {Kindler}, \citenamefont {Lommel},
  \citenamefont {Mann}, \citenamefont {Nishio}, \citenamefont {Page},
  \citenamefont {Ressler}, \citenamefont {Streicher}, \citenamefont
  {\v{S}\'{a}ro}, \citenamefont {Sulignano},\ and\ \citenamefont
  {Van~Duppen}}]{Andreyev2013}%
  \BibitemOpen
  \bibfield  {author} {\bibinfo {author} {\bibfnamefont {A.~N.}\ \bibnamefont
  {Andreyev}}, \bibinfo {author} {\bibfnamefont {S.}~\bibnamefont {Antalic}},
  \bibinfo {author} {\bibfnamefont {D.}~\bibnamefont {Ackermann}}, \bibinfo
  {author} {\bibfnamefont {L.}~\bibnamefont {Bianco}}, \bibinfo {author}
  {\bibfnamefont {S.}~\bibnamefont {Franchoo}}, \bibinfo {author}
  {\bibfnamefont {S.}~\bibnamefont {Heinz}}, \bibinfo {author} {\bibfnamefont
  {F.~P.}\ \bibnamefont {Hessberger}}, \bibinfo {author} {\bibfnamefont
  {S.}~\bibnamefont {Hofmann}}, \bibinfo {author} {\bibfnamefont
  {M.}~\bibnamefont {Huyse}}, \bibinfo {author} {\bibfnamefont
  {Z.}~\bibnamefont {Kalaninov\'{a}}}, \bibinfo {author} {\bibfnamefont
  {I.}~\bibnamefont {Kojouharov}}, \bibinfo {author} {\bibfnamefont
  {B.}~\bibnamefont {Kindler}}, \bibinfo {author} {\bibfnamefont
  {B.}~\bibnamefont {Lommel}}, \bibinfo {author} {\bibfnamefont
  {R.}~\bibnamefont {Mann}}, \bibinfo {author} {\bibfnamefont {K.}~\bibnamefont
  {Nishio}}, \bibinfo {author} {\bibfnamefont {R.~D.}\ \bibnamefont {Page}},
  \bibinfo {author} {\bibfnamefont {J.~J.}\ \bibnamefont {Ressler}}, \bibinfo
  {author} {\bibfnamefont {B.}~\bibnamefont {Streicher}}, \bibinfo {author}
  {\bibfnamefont {{\v S}.}~\bibnamefont {\v{S}\'{a}ro}}, \bibinfo {author}
  {\bibfnamefont {B.}~\bibnamefont {Sulignano}}, \ and\ \bibinfo {author}
  {\bibfnamefont {P.}~\bibnamefont {Van~Duppen}},\ }\href {\doibase
  10.1103/PhysRevC.87.014317} {\bibfield  {journal} {\bibinfo  {journal} {Phys.
  Rev. C}\ }\textbf {\bibinfo {volume} {87}},\ \bibinfo {pages} {014317}
  (\bibinfo {year} {2013}{\natexlab{a}})}\BibitemShut {NoStop}%
\bibitem [{\citenamefont {Lane}\ \emph {et~al.}(2013)\citenamefont {Lane},
  \citenamefont {Andreyev}, \citenamefont {Antalic}, \citenamefont {Ackermann},
  \citenamefont {Gerl}, \citenamefont {Hessberger}, \citenamefont {Hofmann},
  \citenamefont {Huyse}, \citenamefont {Kettunen}, \citenamefont
  {Kleinb\"{o}hl}, \citenamefont {Kindler}, \citenamefont {Kojouharov},
  \citenamefont {Leino}, \citenamefont {Lommel}, \citenamefont
  {M\"{u}nzenberg}, \citenamefont {Nishio}, \citenamefont {Page}, \citenamefont
  {\v{S}\'{a}ro}, \citenamefont {Schaffner}, \citenamefont {Taylor},\ and\
  \citenamefont {{Van Duppen}}}]{Lane2013}%
  \BibitemOpen
  \bibfield  {author} {\bibinfo {author} {\bibfnamefont {J.~F.~W.}\
  \bibnamefont {Lane}}, \bibinfo {author} {\bibfnamefont {A.~N.}\ \bibnamefont
  {Andreyev}}, \bibinfo {author} {\bibfnamefont {S.}~\bibnamefont {Antalic}},
  \bibinfo {author} {\bibfnamefont {D.}~\bibnamefont {Ackermann}}, \bibinfo
  {author} {\bibfnamefont {J.}~\bibnamefont {Gerl}}, \bibinfo {author}
  {\bibfnamefont {F.~P.}\ \bibnamefont {Hessberger}}, \bibinfo {author}
  {\bibfnamefont {S.}~\bibnamefont {Hofmann}}, \bibinfo {author} {\bibfnamefont
  {M.}~\bibnamefont {Huyse}}, \bibinfo {author} {\bibfnamefont
  {H.}~\bibnamefont {Kettunen}}, \bibinfo {author} {\bibfnamefont
  {A.}~\bibnamefont {Kleinb\"{o}hl}}, \bibinfo {author} {\bibfnamefont
  {B.}~\bibnamefont {Kindler}}, \bibinfo {author} {\bibfnamefont
  {I.}~\bibnamefont {Kojouharov}}, \bibinfo {author} {\bibfnamefont
  {M.}~\bibnamefont {Leino}}, \bibinfo {author} {\bibfnamefont
  {B.}~\bibnamefont {Lommel}}, \bibinfo {author} {\bibfnamefont
  {G.}~\bibnamefont {M\"{u}nzenberg}}, \bibinfo {author} {\bibfnamefont
  {K.}~\bibnamefont {Nishio}}, \bibinfo {author} {\bibfnamefont {R.~D.}\
  \bibnamefont {Page}}, \bibinfo {author} {\bibfnamefont {{\v S}.}~\bibnamefont
  {\v{S}\'{a}ro}}, \bibinfo {author} {\bibfnamefont {H.}~\bibnamefont
  {Schaffner}}, \bibinfo {author} {\bibfnamefont {M.~J.}\ \bibnamefont
  {Taylor}}, \ and\ \bibinfo {author} {\bibfnamefont {P.}~\bibnamefont {{Van
  Duppen}}},\ }\href {\doibase 10.1103/PhysRevC.87.014318} {\bibfield
  {journal} {\bibinfo  {journal} {Phys. Rev. C}\ }\textbf {\bibinfo {volume}
  {87}},\ \bibinfo {pages} {014318} (\bibinfo {year} {2013})}\BibitemShut
  {NoStop}%
\bibitem [{\citenamefont {Andreyev}\ \emph
  {et~al.}(2013{\natexlab{b}})\citenamefont {Andreyev}, \citenamefont {Huyse},\
  and\ \citenamefont {{Van Duppen}}}]{Andreyev2013a}%
  \BibitemOpen
  \bibfield  {author} {\bibinfo {author} {\bibfnamefont {A.~N.}\ \bibnamefont
  {Andreyev}}, \bibinfo {author} {\bibfnamefont {M.}~\bibnamefont {Huyse}}, \
  and\ \bibinfo {author} {\bibfnamefont {P.}~\bibnamefont {{Van Duppen}}},\
  }\href {\doibase 10.1103/RevModPhys.85.1541} {\bibfield  {journal} {\bibinfo
  {journal} {Rev. Mod. Phys.}\ }\textbf {\bibinfo {volume} {85}},\ \bibinfo
  {pages} {1541} (\bibinfo {year} {2013}{\natexlab{b}})}\BibitemShut {NoStop}%
\bibitem [{\citenamefont {Panov}\ \emph {et~al.}(2005)\citenamefont {Panov},
  \citenamefont {Kolbe}, \citenamefont {Pfeiffer}, \citenamefont {Rauscher},
  \citenamefont {Kratz},\ and\ \citenamefont {Thielemann}}]{Panov2005}%
  \BibitemOpen
  \bibfield  {author} {\bibinfo {author} {\bibfnamefont {I.~V.}\ \bibnamefont
  {Panov}}, \bibinfo {author} {\bibfnamefont {E.}~\bibnamefont {Kolbe}},
  \bibinfo {author} {\bibfnamefont {B.}~\bibnamefont {Pfeiffer}}, \bibinfo
  {author} {\bibfnamefont {T.}~\bibnamefont {Rauscher}}, \bibinfo {author}
  {\bibfnamefont {K.-L.}\ \bibnamefont {Kratz}}, \ and\ \bibinfo {author}
  {\bibfnamefont {F.-K.}\ \bibnamefont {Thielemann}},\ }\href {\doibase
  10.1016/j.nuclphysa.2004.09.115} {\bibfield  {journal} {\bibinfo  {journal}
  {Nucl. Phys. A}\ }\textbf {\bibinfo {volume} {747}},\ \bibinfo {pages} {633}
  (\bibinfo {year} {2005})}\BibitemShut {NoStop}%
\bibitem [{\citenamefont {Petermann}\ \emph {et~al.}(2012)\citenamefont
  {Petermann}, \citenamefont {Langanke}, \citenamefont {Mart\'{\i}nez-Pinedo},
  \citenamefont {Panov}, \citenamefont {Reinhard},\ and\ \citenamefont
  {Thielemann}}]{Petermann2012}%
  \BibitemOpen
  \bibfield  {author} {\bibinfo {author} {\bibfnamefont {I.}~\bibnamefont
  {Petermann}}, \bibinfo {author} {\bibfnamefont {K.}~\bibnamefont {Langanke}},
  \bibinfo {author} {\bibfnamefont {G.}~\bibnamefont {Mart\'{\i}nez-Pinedo}},
  \bibinfo {author} {\bibfnamefont {I.~V.}\ \bibnamefont {Panov}}, \bibinfo
  {author} {\bibfnamefont {P.~G.}\ \bibnamefont {Reinhard}}, \ and\ \bibinfo
  {author} {\bibfnamefont {F.~K.}\ \bibnamefont {Thielemann}},\ }\href
  {\doibase 10.1140/epja/i2012-12122-6} {\bibfield  {journal} {\bibinfo
  {journal} {Eur. Phys. J. A}\ }\textbf {\bibinfo {volume} {48}},\ \bibinfo
  {pages} {122} (\bibinfo {year} {2012})}\BibitemShut {NoStop}%
\bibitem [{\citenamefont {Shaughnessy}\ \emph {et~al.}(2000)\citenamefont
  {Shaughnessy}, \citenamefont {Adams}, \citenamefont {Gregorich},
  \citenamefont {Lane}, \citenamefont {Laue}, \citenamefont {Lee},
  \citenamefont {McGrath}, \citenamefont {Patin}, \citenamefont {Strellis},
  \citenamefont {Sylwester}, \citenamefont {Wilk},\ and\ \citenamefont
  {Hoffman}}]{Shaughnessy2000}%
  \BibitemOpen
  \bibfield  {author} {\bibinfo {author} {\bibfnamefont {D.~A.}\ \bibnamefont
  {Shaughnessy}}, \bibinfo {author} {\bibfnamefont {J.~L.}\ \bibnamefont
  {Adams}}, \bibinfo {author} {\bibfnamefont {K.~E.}\ \bibnamefont
  {Gregorich}}, \bibinfo {author} {\bibfnamefont {M.~R.}\ \bibnamefont {Lane}},
  \bibinfo {author} {\bibfnamefont {C.~A.}\ \bibnamefont {Laue}}, \bibinfo
  {author} {\bibfnamefont {D.~M.}\ \bibnamefont {Lee}}, \bibinfo {author}
  {\bibfnamefont {C.~A.}\ \bibnamefont {McGrath}}, \bibinfo {author}
  {\bibfnamefont {J.~B.}\ \bibnamefont {Patin}}, \bibinfo {author}
  {\bibfnamefont {D.~A.}\ \bibnamefont {Strellis}}, \bibinfo {author}
  {\bibfnamefont {E.~R.}\ \bibnamefont {Sylwester}}, \bibinfo {author}
  {\bibfnamefont {P.~A.}\ \bibnamefont {Wilk}}, \ and\ \bibinfo {author}
  {\bibfnamefont {D.~C.}\ \bibnamefont {Hoffman}},\ }\href
  {http://www.ncbi.nlm.nih.gov/pubmed/9969915
  http://journals.aps.org/prc/abstract/10.1103/PhysRevC.61.044609} {\bibfield
  {journal} {\bibinfo  {journal} {Phys. Rev. C}\ }\textbf {\bibinfo {volume}
  {61}},\ \bibinfo {pages} {044609} (\bibinfo {year} {2000})}\BibitemShut
  {NoStop}%
\bibitem [{\citenamefont {Shaughnessy}\ \emph {et~al.}(2002)\citenamefont
  {Shaughnessy}, \citenamefont {Gregorich}, \citenamefont {Adams},
  \citenamefont {Lane}, \citenamefont {Laue}, \citenamefont {Lee},
  \citenamefont {McGrath}, \citenamefont {Ninov}, \citenamefont {Patin},
  \citenamefont {Strellis}, \citenamefont {Sylwester}, \citenamefont {Wilk},\
  and\ \citenamefont {Hoffman}}]{Shaughnessy2002}%
  \BibitemOpen
  \bibfield  {author} {\bibinfo {author} {\bibfnamefont {D.~A.}\ \bibnamefont
  {Shaughnessy}}, \bibinfo {author} {\bibfnamefont {K.~E.}\ \bibnamefont
  {Gregorich}}, \bibinfo {author} {\bibfnamefont {J.~L.}\ \bibnamefont
  {Adams}}, \bibinfo {author} {\bibfnamefont {M.~R.}\ \bibnamefont {Lane}},
  \bibinfo {author} {\bibfnamefont {C.~A.}\ \bibnamefont {Laue}}, \bibinfo
  {author} {\bibfnamefont {D.~M.}\ \bibnamefont {Lee}}, \bibinfo {author}
  {\bibfnamefont {C.~A.}\ \bibnamefont {McGrath}}, \bibinfo {author}
  {\bibfnamefont {V.}~\bibnamefont {Ninov}}, \bibinfo {author} {\bibfnamefont
  {J.~B.}\ \bibnamefont {Patin}}, \bibinfo {author} {\bibfnamefont {D.~A.}\
  \bibnamefont {Strellis}}, \bibinfo {author} {\bibfnamefont {E.~R.}\
  \bibnamefont {Sylwester}}, \bibinfo {author} {\bibfnamefont {P.~A.}\
  \bibnamefont {Wilk}}, \ and\ \bibinfo {author} {\bibfnamefont {D.~C.}\
  \bibnamefont {Hoffman}},\ }\href {\doibase 10.1103/PhysRevC.65.024612}
  {\bibfield  {journal} {\bibinfo  {journal} {Phys. Rev. C}\ }\textbf {\bibinfo
  {volume} {65}},\ \bibinfo {pages} {024612} (\bibinfo {year}
  {2002})}\BibitemShut {NoStop}%
\bibitem [{\citenamefont {Britt}\ \emph {et~al.}(1980)\citenamefont {Britt},
  \citenamefont {Cheifetz}, \citenamefont {Hoffman}, \citenamefont {Wilhelmy},
  \citenamefont {Dupzyk},\ and\ \citenamefont {Lougheed}}]{Britt1980}%
  \BibitemOpen
  \bibfield  {author} {\bibinfo {author} {\bibfnamefont {H.~C.}\ \bibnamefont
  {Britt}}, \bibinfo {author} {\bibfnamefont {E.}~\bibnamefont {Cheifetz}},
  \bibinfo {author} {\bibfnamefont {D.~C.}\ \bibnamefont {Hoffman}}, \bibinfo
  {author} {\bibfnamefont {J.~B.}\ \bibnamefont {Wilhelmy}}, \bibinfo {author}
  {\bibfnamefont {R.~J.}\ \bibnamefont {Dupzyk}}, \ and\ \bibinfo {author}
  {\bibfnamefont {R.~W.}\ \bibnamefont {Lougheed}},\ }\href {\doibase
  10.1103/PhysRevC.21.761} {\bibfield  {journal} {\bibinfo  {journal} {Phys.
  Rev. C}\ }\textbf {\bibinfo {volume} {21}},\ \bibinfo {pages} {761} (\bibinfo
  {year} {1980})}\BibitemShut {NoStop}%
\bibitem [{\citenamefont {Audi}\ \emph {et~al.}(2012)\citenamefont {Audi},
  \citenamefont {Kondev}, \citenamefont {Wang}, \citenamefont {Pfeiffer},
  \citenamefont {Sun}, \citenamefont {Blachot},\ and\ \citenamefont
  {MacCormick}}]{Nubase2012}%
  \BibitemOpen
  \bibfield  {author} {\bibinfo {author} {\bibfnamefont {G.}~\bibnamefont
  {Audi}}, \bibinfo {author} {\bibfnamefont {F.~G.~.}\ \bibnamefont {Kondev}},
  \bibinfo {author} {\bibfnamefont {M.}~\bibnamefont {Wang}}, \bibinfo {author}
  {\bibfnamefont {B.}~\bibnamefont {Pfeiffer}}, \bibinfo {author}
  {\bibfnamefont {X.}~\bibnamefont {Sun}}, \bibinfo {author} {\bibfnamefont
  {J.}~\bibnamefont {Blachot}}, \ and\ \bibinfo {author} {\bibfnamefont
  {M.}~\bibnamefont {MacCormick}},\ }\href
  {http://iopscience.iop.org/1674-1137/36/12/001} {\bibfield  {journal}
  {\bibinfo  {journal} {Chinese Phys. C}\ }\textbf {\bibinfo {volume} {36}},\
  \bibinfo {pages} {1157} (\bibinfo {year} {2012})}\BibitemShut {NoStop}%
\bibitem [{\citenamefont {Gangrsky}\ \emph {et~al.}(1980)\citenamefont
  {Gangrsky}, \citenamefont {Miller}, \citenamefont {Mikhailov},\ and\
  \citenamefont {Kharisov}}]{Gangrsky1980}%
  \BibitemOpen
  \bibfield  {author} {\bibinfo {author} {\bibfnamefont {Y.~P.}\ \bibnamefont
  {Gangrsky}}, \bibinfo {author} {\bibfnamefont {M.~B.}\ \bibnamefont
  {Miller}}, \bibinfo {author} {\bibfnamefont {L.~V.}\ \bibnamefont
  {Mikhailov}}, \ and\ \bibinfo {author} {\bibfnamefont {I.~F.}\ \bibnamefont
  {Kharisov}},\ }\href@noop {} {\bibfield  {journal} {\bibinfo  {journal} {Sov.
  J. Nucl. Phys.}\ }\textbf {\bibinfo {volume} {31}},\ \bibinfo {pages} {162}
  (\bibinfo {year} {1980})}\BibitemShut {NoStop}%
\bibitem [{\citenamefont {Klapdor}(1983)}]{Klapdor1983}%
  \BibitemOpen
  \bibfield  {author} {\bibinfo {author} {\bibfnamefont {H.~V.}\ \bibnamefont
  {Klapdor}},\ }\href {\doibase 10.1016/0146-6410(83)90004-2} {\bibfield
  {journal} {\bibinfo  {journal} {Prog. Part. Nucl. Phys.}\ }\textbf {\bibinfo
  {volume} {10}},\ \bibinfo {pages} {131} (\bibinfo {year} {1983})}\BibitemShut
  {NoStop}%
\bibitem [{\citenamefont {Kuznetsov}\ and\ \citenamefont
  {Skobelev}(1999)}]{Kuznetsov1999}%
  \BibitemOpen
  \bibfield  {author} {\bibinfo {author} {\bibfnamefont {V.~I.}\ \bibnamefont
  {Kuznetsov}}\ and\ \bibinfo {author} {\bibfnamefont {N.~K.}\ \bibnamefont
  {Skobelev}},\ }\href {\doibase 10.1134/1.953123} {\bibfield  {journal}
  {\bibinfo  {journal} {Phys. Part. Nucl.}\ }\textbf {\bibinfo {volume} {30}},\
  \bibinfo {pages} {666} (\bibinfo {year} {1999})}\BibitemShut {NoStop}%
\bibitem [{\citenamefont {Kratz}\ and\ \citenamefont
  {Herrmann}(1973)}]{Kratz1973}%
  \BibitemOpen
  \bibfield  {author} {\bibinfo {author} {\bibfnamefont {K.~L.}\ \bibnamefont
  {Kratz}}\ and\ \bibinfo {author} {\bibfnamefont {G.}~\bibnamefont
  {Herrmann}},\ }\href {\doibase 10.1007/BF01391992} {\bibfield  {journal}
  {\bibinfo  {journal} {Z. Phys.}\ }\textbf {\bibinfo {volume} {263}},\
  \bibinfo {pages} {435} (\bibinfo {year} {1973})}\BibitemShut {NoStop}%
\bibitem [{\citenamefont {Hornshoj}\ \emph {et~al.}(1974)\citenamefont
  {Hornshoj}, \citenamefont {Erdal}, \citenamefont {Hansen}, \citenamefont
  {Jonson}, \citenamefont {Aleklett},\ and\ \citenamefont
  {Nyman}}]{Hornshoj1974}%
  \BibitemOpen
  \bibfield  {author} {\bibinfo {author} {\bibfnamefont {P.}~\bibnamefont
  {Hornshoj}}, \bibinfo {author} {\bibfnamefont {B.~R.}\ \bibnamefont {Erdal}},
  \bibinfo {author} {\bibfnamefont {P.~G.}\ \bibnamefont {Hansen}}, \bibinfo
  {author} {\bibfnamefont {B.}~\bibnamefont {Jonson}}, \bibinfo {author}
  {\bibfnamefont {K.}~\bibnamefont {Aleklett}}, \ and\ \bibinfo {author}
  {\bibfnamefont {G.}~\bibnamefont {Nyman}},\ }\href
  {http://www.sciencedirect.com/science/article/pii/0375947475911306}
  {\bibfield  {journal} {\bibinfo  {journal} {Nucl. Phys. A}\ }\textbf
  {\bibinfo {volume} {239}},\ \bibinfo {pages} {15} (\bibinfo {year}
  {1974})}\BibitemShut {NoStop}%
\bibitem [{\citenamefont {Izosimov}\ \emph {et~al.}(2011)\citenamefont
  {Izosimov}, \citenamefont {Kalinnikov},\ and\ \citenamefont
  {Solnyshkin}}]{Izosimov2011}%
  \BibitemOpen
  \bibfield  {author} {\bibinfo {author} {\bibfnamefont {I.~N.}\ \bibnamefont
  {Izosimov}}, \bibinfo {author} {\bibfnamefont {V.~G.}\ \bibnamefont
  {Kalinnikov}}, \ and\ \bibinfo {author} {\bibfnamefont {A.~A.}\ \bibnamefont
  {Solnyshkin}},\ }\href {\doibase 10.1134/S1063779611060049} {\bibfield
  {journal} {\bibinfo  {journal} {Phys. Part. Nucl.}\ }\textbf {\bibinfo
  {volume} {42}},\ \bibinfo {pages} {963} (\bibinfo {year} {2011})}\BibitemShut
  {NoStop}%
\bibitem [{\citenamefont {Veselsk\'{y}}\ \emph {et~al.}(2012)\citenamefont
  {Veselsk\'{y}}, \citenamefont {Andreyev}, \citenamefont {Antalic},
  \citenamefont {Huyse}, \citenamefont {M\"{o}ller}, \citenamefont {Nishio},
  \citenamefont {Sierk}, \citenamefont {{Van Duppen}},\ and\ \citenamefont
  {Venhart}}]{Veselsky2012}%
  \BibitemOpen
  \bibfield  {author} {\bibinfo {author} {\bibfnamefont {M.}~\bibnamefont
  {Veselsk\'{y}}}, \bibinfo {author} {\bibfnamefont {A.~N.}\ \bibnamefont
  {Andreyev}}, \bibinfo {author} {\bibfnamefont {S.}~\bibnamefont {Antalic}},
  \bibinfo {author} {\bibfnamefont {M.}~\bibnamefont {Huyse}}, \bibinfo
  {author} {\bibfnamefont {P.}~\bibnamefont {M\"{o}ller}}, \bibinfo {author}
  {\bibfnamefont {K.}~\bibnamefont {Nishio}}, \bibinfo {author} {\bibfnamefont
  {A.~J.}\ \bibnamefont {Sierk}}, \bibinfo {author} {\bibfnamefont
  {P.}~\bibnamefont {{Van Duppen}}}, \ and\ \bibinfo {author} {\bibfnamefont
  {M.}~\bibnamefont {Venhart}},\ }\href {\doibase 10.1103/PhysRevC.86.024308}
  {\bibfield  {journal} {\bibinfo  {journal} {Phys. Rev. C}\ }\textbf {\bibinfo
  {volume} {86}},\ \bibinfo {pages} {024308} (\bibinfo {year}
  {2012})}\BibitemShut {NoStop}%
\bibitem [{\citenamefont {Habs}\ \emph {et~al.}(1978)\citenamefont {Habs},
  \citenamefont {Klewe-Nebenius}, \citenamefont {Metag}, \citenamefont
  {Neumann},\ and\ \citenamefont {Specht}}]{Habs1978}%
  \BibitemOpen
  \bibfield  {author} {\bibinfo {author} {\bibfnamefont {D.}~\bibnamefont
  {Habs}}, \bibinfo {author} {\bibfnamefont {H.}~\bibnamefont
  {Klewe-Nebenius}}, \bibinfo {author} {\bibfnamefont {V.}~\bibnamefont
  {Metag}}, \bibinfo {author} {\bibfnamefont {B.}~\bibnamefont {Neumann}}, \
  and\ \bibinfo {author} {\bibfnamefont {H.~J.}\ \bibnamefont {Specht}},\
  }\href@noop {} {\bibfield  {journal} {\bibinfo  {journal} {Z. Phys.}\
  }\textbf {\bibinfo {volume} {285}},\ \bibinfo {pages} {53} (\bibinfo {year}
  {1978})}\BibitemShut {NoStop}%
\bibitem [{\citenamefont {Hall}\ and\ \citenamefont
  {Hoffman}(1992)}]{Hall1992}%
  \BibitemOpen
  \bibfield  {author} {\bibinfo {author} {\bibfnamefont {H.~L.}\ \bibnamefont
  {Hall}}\ and\ \bibinfo {author} {\bibfnamefont {D.~C.}\ \bibnamefont
  {Hoffman}},\ }\href {\doibase 10.1146/annurev.ns.42.120192.001051} {\bibfield
   {journal} {\bibinfo  {journal} {Annu. Rev. Nucl. Part. Sci.}\ }\textbf
  {\bibinfo {volume} {42}},\ \bibinfo {pages} {147} (\bibinfo {year}
  {1992})}\BibitemShut {NoStop}%
\bibitem [{\citenamefont {Emeric}\ and\ \citenamefont {Sonzogni}()}]{logft}%
  \BibitemOpen
  \bibfield  {author} {\bibinfo {author} {\bibfnamefont {M.}~\bibnamefont
  {Emeric}}\ and\ \bibinfo {author} {\bibfnamefont {A.}~\bibnamefont
  {Sonzogni}},\ }\href {http://www.nndc.bnl.gov/logft/} {\enquote {\bibinfo
  {title} {\url{http://www.nndc.bnl.gov/logft/}},}\ }\BibitemShut {NoStop}%
\bibitem [{\citenamefont {Firestone}(1999)}]{Firestone1999}%
  \BibitemOpen
  \bibfield  {author} {\bibinfo {author} {\bibfnamefont {R.~B.}\ \bibnamefont
  {Firestone}},\ }\href
  {http://eu.wiley.com/WileyCDA/WileyTitle/productCd-0471356336.html} {\emph
  {\bibinfo {title} {{Table of isotopes}}}},\ \bibinfo {edition} {8th}\ ed.,\
  edited by\ \bibinfo {editor} {\bibfnamefont {V.~S.}\ \bibnamefont {Shirley}}\
  (\bibinfo  {publisher} {Wiley},\ \bibinfo {year} {1999})\ pp.\ \bibinfo
  {pages} {14188--14192}\BibitemShut {NoStop}%
\bibitem [{\citenamefont {M\"{o}ller}\ \emph {et~al.}(2012)\citenamefont
  {M\"{o}ller}, \citenamefont {Randrup},\ and\ \citenamefont
  {Sierk}}]{Moller2012}%
  \BibitemOpen
  \bibfield  {author} {\bibinfo {author} {\bibfnamefont {P.}~\bibnamefont
  {M\"{o}ller}}, \bibinfo {author} {\bibfnamefont {J.}~\bibnamefont {Randrup}},
  \ and\ \bibinfo {author} {\bibfnamefont {A.~J.}\ \bibnamefont {Sierk}},\
  }\href {\doibase 10.1103/PhysRevC.85.024306} {\bibfield  {journal} {\bibinfo
  {journal} {Phys. Rev. C}\ }\textbf {\bibinfo {volume} {85}},\ \bibinfo
  {pages} {024306} (\bibinfo {year} {2012})}\BibitemShut {NoStop}%
\bibitem [{\citenamefont {Wang}\ \emph {et~al.}(2012)\citenamefont {Wang},
  \citenamefont {Audi}, \citenamefont {Wapstra}, \citenamefont {Kondev},
  \citenamefont {MacCormick}, \citenamefont {Xu},\ and\ \citenamefont
  {Pfeiffer}}]{AME2012}%
  \BibitemOpen
  \bibfield  {author} {\bibinfo {author} {\bibfnamefont {M.}~\bibnamefont
  {Wang}}, \bibinfo {author} {\bibfnamefont {G.}~\bibnamefont {Audi}}, \bibinfo
  {author} {\bibfnamefont {A.~H.}\ \bibnamefont {Wapstra}}, \bibinfo {author}
  {\bibfnamefont {F.~G.}\ \bibnamefont {Kondev}}, \bibinfo {author}
  {\bibfnamefont {M.}~\bibnamefont {MacCormick}}, \bibinfo {author}
  {\bibfnamefont {X.}~\bibnamefont {Xu}}, \ and\ \bibinfo {author}
  {\bibfnamefont {B.}~\bibnamefont {Pfeiffer}},\ }\href {\doibase
  10.1088/1674-1137/36/12/003} {\bibfield  {journal} {\bibinfo  {journal}
  {Chinese Phys. C}\ }\textbf {\bibinfo {volume} {36}},\ \bibinfo {pages}
  {1603} (\bibinfo {year} {2012})}\BibitemShut {NoStop}%
\bibitem [{\citenamefont {Zagrebaev}\ \emph {et~al.}()\citenamefont
  {Zagrebaev}, \citenamefont {Denikin}, \citenamefont {Alekseev}, \citenamefont
  {Karpov}, \citenamefont {Samarin}, \citenamefont {Naumenko},\ and\
  \citenamefont {Kozhin}}]{NRV}%
  \BibitemOpen
  \bibfield  {author} {\bibinfo {author} {\bibfnamefont {V.}~\bibnamefont
  {Zagrebaev}}, \bibinfo {author} {\bibfnamefont {A.}~\bibnamefont {Denikin}},
  \bibinfo {author} {\bibfnamefont {A.~P.}\ \bibnamefont {Alekseev}}, \bibinfo
  {author} {\bibfnamefont {A.~V.}\ \bibnamefont {Karpov}}, \bibinfo {author}
  {\bibfnamefont {V.~V.}\ \bibnamefont {Samarin}}, \bibinfo {author}
  {\bibfnamefont {M.~A.}\ \bibnamefont {Naumenko}}, \ and\ \bibinfo {author}
  {\bibfnamefont {A.~Y.}\ \bibnamefont {Kozhin}},\ }\href
  {http://nrv.jinr.ru/nrv/} {\enquote {\bibinfo {title}
  {\url{http://nrv.jinr.ru/nrv/}},}\ }\BibitemShut {NoStop}%
\bibitem [{\citenamefont {Mamdouh}\ \emph {et~al.}(2001)\citenamefont
  {Mamdouh}, \citenamefont {Pearson}, \citenamefont {Rayet},\ and\
  \citenamefont {Tondeur}}]{Mamdouh2001}%
  \BibitemOpen
  \bibfield  {author} {\bibinfo {author} {\bibfnamefont {A.}~\bibnamefont
  {Mamdouh}}, \bibinfo {author} {\bibfnamefont {J.~M.}\ \bibnamefont
  {Pearson}}, \bibinfo {author} {\bibfnamefont {M.}~\bibnamefont {Rayet}}, \
  and\ \bibinfo {author} {\bibfnamefont {F.}~\bibnamefont {Tondeur}},\ }\href
  {\doibase 10.1016/S0375-9474(00)00358-4} {\bibfield  {journal} {\bibinfo
  {journal} {Nucl. Phys. A}\ }\textbf {\bibinfo {volume} {679}},\ \bibinfo
  {pages} {337} (\bibinfo {year} {2001})}\BibitemShut {NoStop}%
\bibitem [{\citenamefont {M\"{o}ller}\ \emph {et~al.}(1995)\citenamefont
  {M\"{o}ller}, \citenamefont {Nix}, \citenamefont {Myers},\ and\ \citenamefont
  {Swiatecki}}]{Moller1995}%
  \BibitemOpen
  \bibfield  {author} {\bibinfo {author} {\bibfnamefont {P.}~\bibnamefont
  {M\"{o}ller}}, \bibinfo {author} {\bibfnamefont {J.~R.}\ \bibnamefont {Nix}},
  \bibinfo {author} {\bibfnamefont {W.~D.}\ \bibnamefont {Myers}}, \ and\
  \bibinfo {author} {\bibfnamefont {W.~J.}\ \bibnamefont {Swiatecki}},\ }\href
  {http://www.sciencedirect.com/science/article/pii/S0092640X85710029}
  {\bibfield  {journal} {\bibinfo  {journal} {At. Data Nucl. Data Tables}\
  }\textbf {\bibinfo {volume} {59}},\ \bibinfo {pages} {185} (\bibinfo {year}
  {1995})}\BibitemShut {NoStop}%
\bibitem [{\citenamefont {Myers}\ and\ \citenamefont
  {\'{S}wia̧tecki}(1999)}]{Myers1999}%
  \BibitemOpen
  \bibfield  {author} {\bibinfo {author} {\bibfnamefont {W.~D.}\ \bibnamefont
  {Myers}}\ and\ \bibinfo {author} {\bibfnamefont {W.~J.}\ \bibnamefont
  {\'{S}wia̧tecki}},\ }\href {\doibase 10.1103/PhysRevC.60.014606} {\bibfield
  {journal} {\bibinfo  {journal} {Phys. Rev. C}\ }\textbf {\bibinfo {volume}
  {60}},\ \bibinfo {pages} {014606} (\bibinfo {year} {1999})}\BibitemShut
  {NoStop}%
\bibitem [{\citenamefont {M\"oller}\ \emph {et~al.}(2015)\citenamefont
  {M\"oller}, \citenamefont {Sierk}, \citenamefont {Ichikawa}, \citenamefont
  {Iwamoto},\ and\ \citenamefont {Mumpower}}]{Moller2015}%
  \BibitemOpen
  \bibfield  {author} {\bibinfo {author} {\bibfnamefont {P.}~\bibnamefont
  {M\"oller}}, \bibinfo {author} {\bibfnamefont {A.~J.}\ \bibnamefont {Sierk}},
  \bibinfo {author} {\bibfnamefont {T.}~\bibnamefont {Ichikawa}}, \bibinfo
  {author} {\bibfnamefont {A.}~\bibnamefont {Iwamoto}}, \ and\ \bibinfo
  {author} {\bibfnamefont {M.}~\bibnamefont {Mumpower}},\ }\href {\doibase
  10.1103/PhysRevC.91.024310} {\bibfield  {journal} {\bibinfo  {journal} {Phys.
  Rev. C}\ }\textbf {\bibinfo {volume} {91}},\ \bibinfo {pages} {024310}
  (\bibinfo {year} {2015})}\BibitemShut {NoStop}%
\bibitem [{\citenamefont {Myers}\ and\ \citenamefont
  {Swiatecki}(1996)}]{Myers1996}%
  \BibitemOpen
  \bibfield  {author} {\bibinfo {author} {\bibfnamefont {W.~D.}\ \bibnamefont
  {Myers}}\ and\ \bibinfo {author} {\bibfnamefont {W.~J.}\ \bibnamefont
  {Swiatecki}},\ }\href {\doibase 10.1016/0375-9474(95)00509-9} {\bibfield
  {journal} {\bibinfo  {journal} {Nucl. Phys. A}\ }\textbf {\bibinfo {volume}
  {601}},\ \bibinfo {pages} {141} (\bibinfo {year} {1996})}\BibitemShut
  {NoStop}%
\bibitem [{\citenamefont {M\"{o}ller}\ \emph {et~al.}(1997)\citenamefont
  {M\"{o}ller}, \citenamefont {Nix},\ and\ \citenamefont {Kratz}}]{Moller1997}%
  \BibitemOpen
  \bibfield  {author} {\bibinfo {author} {\bibfnamefont {P.}~\bibnamefont
  {M\"{o}ller}}, \bibinfo {author} {\bibfnamefont {J.~R.}\ \bibnamefont {Nix}},
  \ and\ \bibinfo {author} {\bibfnamefont {K.-L.}\ \bibnamefont {Kratz}},\
  }\href {\doibase 10.1006/adnd.1997.0746} {\bibfield  {journal} {\bibinfo
  {journal} {At. Data Nucl. Data Tables}\ }\textbf {\bibinfo {volume} {66}},\
  \bibinfo {pages} {131} (\bibinfo {year} {1997})}\BibitemShut {NoStop}%
\bibitem [{\citenamefont {{Truesdale et al. \textit{(in
  preparation)}}}(2015)}]{Truesdale2014}%
  \BibitemOpen
  \bibfield  {author} {\bibinfo {author} {\bibfnamefont {V.~L.}\ \bibnamefont
  {{Truesdale et al. \textit{(in preparation)}}}},\ }\href@noop {} {\
  (\bibinfo {year} {2015})}\BibitemShut {NoStop}%
\bibitem [{\citenamefont {Kalaninov\'{a}}\ \emph {et~al.}(2014)\citenamefont
  {Kalaninov\'{a}}, \citenamefont {Antalic}, \citenamefont {Andreyev},
  \citenamefont {Hessberger}, \citenamefont {Ackermann}, \citenamefont {Andel},
  \citenamefont {Bianco}, \citenamefont {Hofmann}, \citenamefont {Huyse},
  \citenamefont {Kindler}, \citenamefont {Lommel}, \citenamefont {Mann},
  \citenamefont {Page}, \citenamefont {Sapple}, \citenamefont {Thomson},
  \citenamefont {{Van Duppen}},\ and\ \citenamefont
  {Venhart}}]{Kalaninova2014}%
  \BibitemOpen
  \bibfield  {author} {\bibinfo {author} {\bibfnamefont {Z.}~\bibnamefont
  {Kalaninov\'{a}}}, \bibinfo {author} {\bibfnamefont {S.}~\bibnamefont
  {Antalic}}, \bibinfo {author} {\bibfnamefont {A.~N.}\ \bibnamefont
  {Andreyev}}, \bibinfo {author} {\bibfnamefont {F.~P.}\ \bibnamefont
  {Hessberger}}, \bibinfo {author} {\bibfnamefont {D.}~\bibnamefont
  {Ackermann}}, \bibinfo {author} {\bibfnamefont {B.}~\bibnamefont {Andel}},
  \bibinfo {author} {\bibfnamefont {L.}~\bibnamefont {Bianco}}, \bibinfo
  {author} {\bibfnamefont {S.}~\bibnamefont {Hofmann}}, \bibinfo {author}
  {\bibfnamefont {M.}~\bibnamefont {Huyse}}, \bibinfo {author} {\bibfnamefont
  {B.}~\bibnamefont {Kindler}}, \bibinfo {author} {\bibfnamefont
  {B.}~\bibnamefont {Lommel}}, \bibinfo {author} {\bibfnamefont
  {R.}~\bibnamefont {Mann}}, \bibinfo {author} {\bibfnamefont {R.~D.}\
  \bibnamefont {Page}}, \bibinfo {author} {\bibfnamefont {P.~J.}\ \bibnamefont
  {Sapple}}, \bibinfo {author} {\bibfnamefont {J.}~\bibnamefont {Thomson}},
  \bibinfo {author} {\bibfnamefont {P.}~\bibnamefont {{Van Duppen}}}, \ and\
  \bibinfo {author} {\bibfnamefont {M.}~\bibnamefont {Venhart}},\ }\href
  {\doibase 10.1103/PhysRevC.89.054312} {\bibfield  {journal} {\bibinfo
  {journal} {Phys. Rev. C}\ }\textbf {\bibinfo {volume} {89}},\ \bibinfo
  {pages} {054312} (\bibinfo {year} {2014})}\BibitemShut {NoStop}%
\bibitem [{\citenamefont {Kreek}\ \emph
  {et~al.}(1994{\natexlab{a}})\citenamefont {Kreek}, \citenamefont {Hall},
  \citenamefont {Gregorich}, \citenamefont {Henderson}, \citenamefont {Leyba},
  \citenamefont {Czerwinski}, \citenamefont {Kadkhodayan}, \citenamefont {Neu},
  \citenamefont {Kacher}, \citenamefont {Hamilton}, \citenamefont {Lane},
  \citenamefont {Sylwester}, \citenamefont {T\"{u}rler}, \citenamefont {Lee},
  \citenamefont {Nurmia},\ and\ \citenamefont {Hoffman}}]{Kreek1994}%
  \BibitemOpen
  \bibfield  {author} {\bibinfo {author} {\bibfnamefont {S.~A.}\ \bibnamefont
  {Kreek}}, \bibinfo {author} {\bibfnamefont {H.~L.}\ \bibnamefont {Hall}},
  \bibinfo {author} {\bibfnamefont {K.~E.}\ \bibnamefont {Gregorich}}, \bibinfo
  {author} {\bibfnamefont {R.~A.}\ \bibnamefont {Henderson}}, \bibinfo {author}
  {\bibfnamefont {J.~D.}\ \bibnamefont {Leyba}}, \bibinfo {author}
  {\bibfnamefont {K.~R.}\ \bibnamefont {Czerwinski}}, \bibinfo {author}
  {\bibfnamefont {B.}~\bibnamefont {Kadkhodayan}}, \bibinfo {author}
  {\bibfnamefont {M.~P.}\ \bibnamefont {Neu}}, \bibinfo {author} {\bibfnamefont
  {C.~D.}\ \bibnamefont {Kacher}}, \bibinfo {author} {\bibfnamefont {T.~M.}\
  \bibnamefont {Hamilton}}, \bibinfo {author} {\bibfnamefont {M.~R.}\
  \bibnamefont {Lane}}, \bibinfo {author} {\bibfnamefont {E.~R.}\ \bibnamefont
  {Sylwester}}, \bibinfo {author} {\bibfnamefont {A.}~\bibnamefont
  {T\"{u}rler}}, \bibinfo {author} {\bibfnamefont {D.~M.}\ \bibnamefont {Lee}},
  \bibinfo {author} {\bibfnamefont {M.~J.}\ \bibnamefont {Nurmia}}, \ and\
  \bibinfo {author} {\bibfnamefont {D.~C.}\ \bibnamefont {Hoffman}},\ }\href
  {\doibase 10.1103/PhysRevC.50.2288} {\bibfield  {journal} {\bibinfo
  {journal} {Phys. Rev. C}\ }\textbf {\bibinfo {volume} {50}},\ \bibinfo
  {pages} {2288} (\bibinfo {year} {1994}{\natexlab{a}})}\BibitemShut {NoStop}%
\bibitem [{\citenamefont {Hall}\ \emph
  {et~al.}(1990{\natexlab{a}})\citenamefont {Hall}, \citenamefont {Gregorich},
  \citenamefont {Henderson}, \citenamefont {Gannett}, \citenamefont {Chadwick},
  \citenamefont {Leyba}, \citenamefont {Czerwinski}, \citenamefont
  {Kadkhodayan}, \citenamefont {Kreek}, \citenamefont {Hannink}, \citenamefont
  {Lee}, \citenamefont {Nurmia}, \citenamefont {Hoffman}, \citenamefont
  {Palmer},\ and\ \citenamefont {Baisden}}]{Hall1990}%
  \BibitemOpen
  \bibfield  {author} {\bibinfo {author} {\bibfnamefont {H.~L.}\ \bibnamefont
  {Hall}}, \bibinfo {author} {\bibfnamefont {K.~E.}\ \bibnamefont {Gregorich}},
  \bibinfo {author} {\bibfnamefont {R.~A.}\ \bibnamefont {Henderson}}, \bibinfo
  {author} {\bibfnamefont {C.~M.}\ \bibnamefont {Gannett}}, \bibinfo {author}
  {\bibfnamefont {R.~B.}\ \bibnamefont {Chadwick}}, \bibinfo {author}
  {\bibfnamefont {J.~D.}\ \bibnamefont {Leyba}}, \bibinfo {author}
  {\bibfnamefont {K.~R.}\ \bibnamefont {Czerwinski}}, \bibinfo {author}
  {\bibfnamefont {B.}~\bibnamefont {Kadkhodayan}}, \bibinfo {author}
  {\bibfnamefont {S.~A.}\ \bibnamefont {Kreek}}, \bibinfo {author}
  {\bibfnamefont {N.~J.}\ \bibnamefont {Hannink}}, \bibinfo {author}
  {\bibfnamefont {D.~M.}\ \bibnamefont {Lee}}, \bibinfo {author} {\bibfnamefont
  {M.~J.}\ \bibnamefont {Nurmia}}, \bibinfo {author} {\bibfnamefont {D.~C.}\
  \bibnamefont {Hoffman}}, \bibinfo {author} {\bibfnamefont {C.~E.~A.}\
  \bibnamefont {Palmer}}, \ and\ \bibinfo {author} {\bibfnamefont {P.~A.}\
  \bibnamefont {Baisden}},\ }\href {\doibase 10.1103/PhysRevC.42.1480}
  {\bibfield  {journal} {\bibinfo  {journal} {Phys. Rev. C}\ }\textbf {\bibinfo
  {volume} {42}},\ \bibinfo {pages} {1480} (\bibinfo {year}
  {1990}{\natexlab{a}})}\BibitemShut {NoStop}%
\bibitem [{\citenamefont {Hall}\ \emph
  {et~al.}(1990{\natexlab{b}})\citenamefont {Hall}, \citenamefont {Gregorich},
  \citenamefont {Henderson}, \citenamefont {Gannett}, \citenamefont {Chadwick},
  \citenamefont {Leyba}, \citenamefont {Czerwinski}, \citenamefont
  {Kadkhodayan}, \citenamefont {Kreek}, \citenamefont {Lee}, \citenamefont
  {Nurmia}, \citenamefont {Hoffman}, \citenamefont {Palmer},\ and\
  \citenamefont {Baisden}}]{Hall1990a}%
  \BibitemOpen
  \bibfield  {author} {\bibinfo {author} {\bibfnamefont {H.~L.}\ \bibnamefont
  {Hall}}, \bibinfo {author} {\bibfnamefont {K.~E.}\ \bibnamefont {Gregorich}},
  \bibinfo {author} {\bibfnamefont {R.~A.}\ \bibnamefont {Henderson}}, \bibinfo
  {author} {\bibfnamefont {C.~M.}\ \bibnamefont {Gannett}}, \bibinfo {author}
  {\bibfnamefont {R.~B.}\ \bibnamefont {Chadwick}}, \bibinfo {author}
  {\bibfnamefont {J.~D.}\ \bibnamefont {Leyba}}, \bibinfo {author}
  {\bibfnamefont {K.~R.}\ \bibnamefont {Czerwinski}}, \bibinfo {author}
  {\bibfnamefont {B.}~\bibnamefont {Kadkhodayan}}, \bibinfo {author}
  {\bibfnamefont {S.~A.}\ \bibnamefont {Kreek}}, \bibinfo {author}
  {\bibfnamefont {D.~M.}\ \bibnamefont {Lee}}, \bibinfo {author} {\bibfnamefont
  {M.~J.}\ \bibnamefont {Nurmia}}, \bibinfo {author} {\bibfnamefont {D.~C.}\
  \bibnamefont {Hoffman}}, \bibinfo {author} {\bibfnamefont {C.~E.~A.}\
  \bibnamefont {Palmer}}, \ and\ \bibinfo {author} {\bibfnamefont {P.~A.}\
  \bibnamefont {Baisden}},\ }\href
  {http://journals.aps.org/prc/abstract/10.1103/PhysRevC.41.618} {\bibfield
  {journal} {\bibinfo  {journal} {Phys. Rev. C}\ }\textbf {\bibinfo {volume}
  {41}},\ \bibinfo {pages} {618} (\bibinfo {year}
  {1990}{\natexlab{b}})}\BibitemShut {NoStop}%
\bibitem [{\citenamefont {Kreek}\ \emph
  {et~al.}(1994{\natexlab{b}})\citenamefont {Kreek}, \citenamefont {Hall},
  \citenamefont {Gregorich}, \citenamefont {Henderson}, \citenamefont {Leyba},
  \citenamefont {Czerwinski}, \citenamefont {Kadkhodayan}, \citenamefont {Neu},
  \citenamefont {Kacher}, \citenamefont {Hamilton}, \citenamefont {Lane},
  \citenamefont {Sylwester}, \citenamefont {T\"{u}rler}, \citenamefont {Lee},
  \citenamefont {Nurmia},\ and\ \citenamefont {Hoffman}}]{Kreek1994a}%
  \BibitemOpen
  \bibfield  {author} {\bibinfo {author} {\bibfnamefont {S.~A.}\ \bibnamefont
  {Kreek}}, \bibinfo {author} {\bibfnamefont {H.~L.}\ \bibnamefont {Hall}},
  \bibinfo {author} {\bibfnamefont {K.~E.}\ \bibnamefont {Gregorich}}, \bibinfo
  {author} {\bibfnamefont {R.~A.}\ \bibnamefont {Henderson}}, \bibinfo {author}
  {\bibfnamefont {J.~D.}\ \bibnamefont {Leyba}}, \bibinfo {author}
  {\bibfnamefont {K.~R.}\ \bibnamefont {Czerwinski}}, \bibinfo {author}
  {\bibfnamefont {B.}~\bibnamefont {Kadkhodayan}}, \bibinfo {author}
  {\bibfnamefont {M.~P.}\ \bibnamefont {Neu}}, \bibinfo {author} {\bibfnamefont
  {C.~D.}\ \bibnamefont {Kacher}}, \bibinfo {author} {\bibfnamefont {T.~M.}\
  \bibnamefont {Hamilton}}, \bibinfo {author} {\bibfnamefont {M.~R.}\
  \bibnamefont {Lane}}, \bibinfo {author} {\bibfnamefont {E.~R.}\ \bibnamefont
  {Sylwester}}, \bibinfo {author} {\bibfnamefont {A.}~\bibnamefont
  {T\"{u}rler}}, \bibinfo {author} {\bibfnamefont {D.~M.}\ \bibnamefont {Lee}},
  \bibinfo {author} {\bibfnamefont {M.~J.}\ \bibnamefont {Nurmia}}, \ and\
  \bibinfo {author} {\bibfnamefont {D.~C.}\ \bibnamefont {Hoffman}},\ }\href
  {http://journals.aps.org/prc/abstract/10.1103/PhysRevC.49.1859} {\bibfield
  {journal} {\bibinfo  {journal} {Phys. Rev. C}\ }\textbf {\bibinfo {volume}
  {49}} (\bibinfo {year} {1994}{\natexlab{b}})}\BibitemShut {NoStop}%
\bibitem [{\citenamefont {Galeriu}(1983)}]{Galeriu1983}%
  \BibitemOpen
  \bibfield  {author} {\bibinfo {author} {\bibfnamefont {D.}~\bibnamefont
  {Galeriu}},\ }\href {http://iopscience.iop.org/0305-4616/9/3/011} {\bibfield
  {journal} {\bibinfo  {journal} {J. Phys. G Nucl. Phys.}\ }\textbf {\bibinfo
  {volume} {9}},\ \bibinfo {pages} {309} (\bibinfo {year} {1983})}\BibitemShut
  {NoStop}%
\bibitem [{\citenamefont {Antalic}\ \emph {et~al.}(2010)\citenamefont
  {Antalic}, \citenamefont {Hessberger}, \citenamefont {Hofmann}, \citenamefont
  {Ackermann}, \citenamefont {Heinz}, \citenamefont {Kindler}, \citenamefont
  {Kojouharov}, \citenamefont {Kuusiniemi}, \citenamefont {Leino},
  \citenamefont {Lommel}, \citenamefont {Mann},\ and\ \citenamefont
  {\v{S}\'{a}ro}}]{Antalic2009}%
  \BibitemOpen
  \bibfield  {author} {\bibinfo {author} {\bibfnamefont {S.}~\bibnamefont
  {Antalic}}, \bibinfo {author} {\bibfnamefont {F.~P.}\ \bibnamefont
  {Hessberger}}, \bibinfo {author} {\bibfnamefont {S.}~\bibnamefont {Hofmann}},
  \bibinfo {author} {\bibfnamefont {D.}~\bibnamefont {Ackermann}}, \bibinfo
  {author} {\bibfnamefont {S.}~\bibnamefont {Heinz}}, \bibinfo {author}
  {\bibfnamefont {B.}~\bibnamefont {Kindler}}, \bibinfo {author} {\bibfnamefont
  {I.}~\bibnamefont {Kojouharov}}, \bibinfo {author} {\bibfnamefont
  {P.}~\bibnamefont {Kuusiniemi}}, \bibinfo {author} {\bibfnamefont
  {M.}~\bibnamefont {Leino}}, \bibinfo {author} {\bibfnamefont
  {B.}~\bibnamefont {Lommel}}, \bibinfo {author} {\bibfnamefont
  {R.}~\bibnamefont {Mann}}, \ and\ \bibinfo {author} {\bibfnamefont {{\v
  S}.}~\bibnamefont {\v{S}\'{a}ro}},\ }\href {\doibase
  10.1140/epja/i2009-10896-0} {\bibfield  {journal} {\bibinfo  {journal} {Eur.
  Phys. J. A}\ }\textbf {\bibinfo {volume} {43}},\ \bibinfo {pages} {35}
  (\bibinfo {year} {2010})}\BibitemShut {NoStop}%
\bibitem [{\citenamefont {Shaughnessy}\ \emph {et~al.}(2001)\citenamefont
  {Shaughnessy}, \citenamefont {Gregorich}, \citenamefont {Lane}, \citenamefont
  {Laue}, \citenamefont {Lee}, \citenamefont {McGrath}, \citenamefont
  {Strellis}, \citenamefont {Sylwester}, \citenamefont {Wilk},\ and\
  \citenamefont {Hoffman}}]{Shaughnessy2001}%
  \BibitemOpen
  \bibfield  {author} {\bibinfo {author} {\bibfnamefont {D.~A.}\ \bibnamefont
  {Shaughnessy}}, \bibinfo {author} {\bibfnamefont {K.~E.}\ \bibnamefont
  {Gregorich}}, \bibinfo {author} {\bibfnamefont {M.~R.}\ \bibnamefont {Lane}},
  \bibinfo {author} {\bibfnamefont {C.~A.}\ \bibnamefont {Laue}}, \bibinfo
  {author} {\bibfnamefont {D.~M.}\ \bibnamefont {Lee}}, \bibinfo {author}
  {\bibfnamefont {C.~A.}\ \bibnamefont {McGrath}}, \bibinfo {author}
  {\bibfnamefont {D.~A.}\ \bibnamefont {Strellis}}, \bibinfo {author}
  {\bibfnamefont {E.~R.}\ \bibnamefont {Sylwester}}, \bibinfo {author}
  {\bibfnamefont {P.~A.}\ \bibnamefont {Wilk}}, \ and\ \bibinfo {author}
  {\bibfnamefont {D.~C.}\ \bibnamefont {Hoffman}},\ }\href {\doibase
  10.1103/PhysRevC.63.037603} {\bibfield  {journal} {\bibinfo  {journal} {Phys.
  Rev. C}\ }\textbf {\bibinfo {volume} {63}},\ \bibinfo {pages} {037603}
  (\bibinfo {year} {2001})}\BibitemShut {NoStop}%
\bibitem [{\citenamefont {Yanbing}\ \emph {et~al.}(2006)\citenamefont
  {Yanbing}, \citenamefont {Shengdong}, \citenamefont {Huajie}, \citenamefont
  {Shuanggui}, \citenamefont {Weifan}, \citenamefont {Yanning}, \citenamefont
  {Xiting}, \citenamefont {Yingjun},\ and\ \citenamefont
  {Yonghou}}]{Yanbing2006}%
  \BibitemOpen
  \bibfield  {author} {\bibinfo {author} {\bibfnamefont {X.}~\bibnamefont
  {Yanbing}}, \bibinfo {author} {\bibfnamefont {Z.}~\bibnamefont {Shengdong}},
  \bibinfo {author} {\bibfnamefont {D.}~\bibnamefont {Huajie}}, \bibinfo
  {author} {\bibfnamefont {Y.}~\bibnamefont {Shuanggui}}, \bibinfo {author}
  {\bibfnamefont {Y.}~\bibnamefont {Weifan}}, \bibinfo {author} {\bibfnamefont
  {N.}~\bibnamefont {Yanning}}, \bibinfo {author} {\bibfnamefont
  {L.}~\bibnamefont {Xiting}}, \bibinfo {author} {\bibfnamefont
  {L.}~\bibnamefont {Yingjun}}, \ and\ \bibinfo {author} {\bibfnamefont
  {X.}~\bibnamefont {Yonghou}},\ }\href {\doibase 10.1103/PhysRevC.74.047303}
  {\bibfield  {journal} {\bibinfo  {journal} {Phys. Rev. C}\ }\textbf {\bibinfo
  {volume} {74}},\ \bibinfo {pages} {047303} (\bibinfo {year}
  {2006})}\BibitemShut {NoStop}%
\bibitem [{\citenamefont {Shuanggui}\ \emph {et~al.}(2001)\citenamefont
  {Shuanggui}, \citenamefont {Weifan}, \citenamefont {Yanbing}, \citenamefont
  {Qiangyan}, \citenamefont {Bing}, \citenamefont {Jianjun}, \citenamefont
  {Dong}, \citenamefont {Yingjun}, \citenamefont {Taotao},\ and\ \citenamefont
  {Zhenguo}}]{Shuanggui2001}%
  \BibitemOpen
  \bibfield  {author} {\bibinfo {author} {\bibfnamefont {Y.}~\bibnamefont
  {Shuanggui}}, \bibinfo {author} {\bibfnamefont {Y.}~\bibnamefont {Weifan}},
  \bibinfo {author} {\bibfnamefont {X.}~\bibnamefont {Yanbing}}, \bibinfo
  {author} {\bibfnamefont {P.}~\bibnamefont {Qiangyan}}, \bibinfo {author}
  {\bibfnamefont {X.}~\bibnamefont {Bing}}, \bibinfo {author} {\bibfnamefont
  {H.}~\bibnamefont {Jianjun}}, \bibinfo {author} {\bibfnamefont
  {W.}~\bibnamefont {Dong}}, \bibinfo {author} {\bibfnamefont {L.}~\bibnamefont
  {Yingjun}}, \bibinfo {author} {\bibfnamefont {M.}~\bibnamefont {Taotao}}, \
  and\ \bibinfo {author} {\bibfnamefont {Y.}~\bibnamefont {Zhenguo}},\ }\href
  {\doibase 10.1007/s100500170136} {\bibfield  {journal} {\bibinfo  {journal}
  {Eur. Phys. J. A}\ }\textbf {\bibinfo {volume} {10}},\ \bibinfo {pages} {1}
  (\bibinfo {year} {2001})}\BibitemShut {NoStop}%
\bibitem [{\citenamefont {Gangrsky}\ \emph {et~al.}(1978)\citenamefont
  {Gangrsky}, \citenamefont {Marinesky}, \citenamefont {Miller}, \citenamefont
  {Samsuk},\ and\ \citenamefont {Kharisov}}]{Gangrsky1978}%
  \BibitemOpen
  \bibfield  {author} {\bibinfo {author} {\bibfnamefont {Y.~P.}\ \bibnamefont
  {Gangrsky}}, \bibinfo {author} {\bibfnamefont {G.~M.}\ \bibnamefont
  {Marinesky}}, \bibinfo {author} {\bibfnamefont {M.~B.}\ \bibnamefont
  {Miller}}, \bibinfo {author} {\bibfnamefont {V.~N.}\ \bibnamefont {Samsuk}},
  \ and\ \bibinfo {author} {\bibfnamefont {I.~F.}\ \bibnamefont {Kharisov}},\
  }\href@noop {} {\bibfield  {journal} {\bibinfo  {journal} {Sov. J. Nucl.
  Phys.}\ }\textbf {\bibinfo {volume} {27}},\ \bibinfo {pages} {475} (\bibinfo
  {year} {1978})}\BibitemShut {NoStop}%
\bibitem [{\citenamefont {Baas-May}\ \emph {et~al.}(1985)\citenamefont
  {Baas-May}, \citenamefont {Kratz},\ and\ \citenamefont
  {Trautmann}}]{Baas-May1985}%
  \BibitemOpen
  \bibfield  {author} {\bibinfo {author} {\bibfnamefont {A.}~\bibnamefont
  {Baas-May}}, \bibinfo {author} {\bibfnamefont {J.~V.}\ \bibnamefont {Kratz}},
  \ and\ \bibinfo {author} {\bibfnamefont {N.}~\bibnamefont {Trautmann}},\
  }\href {\doibase 10.1007/BF01412080} {\bibfield  {journal} {\bibinfo
  {journal} {Zeitschrift f\"{u}r Phys. A Atoms Nucl.}\ }\textbf {\bibinfo
  {volume} {322}},\ \bibinfo {pages} {457} (\bibinfo {year}
  {1985})}\BibitemShut {NoStop}%
\bibitem [{\citenamefont {Hall}\ \emph {et~al.}(1989)\citenamefont {Hall},
  \citenamefont {Gregorich}, \citenamefont {Henderson}, \citenamefont {Lee},
  \citenamefont {Hoffman}, \citenamefont {Bunker}, \citenamefont {Fowler},
  \citenamefont {Lysaght}, \citenamefont {Starner},\ and\ \citenamefont
  {Wilhelmy}}]{Hall1989}%
  \BibitemOpen
  \bibfield  {author} {\bibinfo {author} {\bibfnamefont {H.~L.}\ \bibnamefont
  {Hall}}, \bibinfo {author} {\bibfnamefont {K.~E.}\ \bibnamefont {Gregorich}},
  \bibinfo {author} {\bibfnamefont {R.~A.}\ \bibnamefont {Henderson}}, \bibinfo
  {author} {\bibfnamefont {D.~M.}\ \bibnamefont {Lee}}, \bibinfo {author}
  {\bibfnamefont {D.~C.}\ \bibnamefont {Hoffman}}, \bibinfo {author}
  {\bibfnamefont {M.~E.}\ \bibnamefont {Bunker}}, \bibinfo {author}
  {\bibfnamefont {M.~M.}\ \bibnamefont {Fowler}}, \bibinfo {author}
  {\bibfnamefont {P.}~\bibnamefont {Lysaght}}, \bibinfo {author} {\bibfnamefont
  {J.~W.}\ \bibnamefont {Starner}}, \ and\ \bibinfo {author} {\bibfnamefont
  {J.~B.}\ \bibnamefont {Wilhelmy}},\ }\href {\doibase
  10.1103/PhysRevC.39.1866} {\bibfield  {journal} {\bibinfo  {journal} {Phys.
  Rev. C}\ }\textbf {\bibinfo {volume} {39}},\ \bibinfo {pages} {1866}
  (\bibinfo {year} {1989})}\BibitemShut {NoStop}%
\bibitem [{\citenamefont {Bohr}\ and\ \citenamefont
  {Wheeler}(1939)}]{Bohr1939}%
  \BibitemOpen
  \bibfield  {author} {\bibinfo {author} {\bibfnamefont {N.}~\bibnamefont
  {Bohr}}\ and\ \bibinfo {author} {\bibfnamefont {J.~A.}\ \bibnamefont
  {Wheeler}},\ }\href
  {http://journals.aps.org/pr/abstract/10.1103/PhysRev.56.426} {\bibfield
  {journal} {\bibinfo  {journal} {Phys. Rev.}\ }\textbf {\bibinfo {volume}
  {56}},\ \bibinfo {pages} {426} (\bibinfo {year} {1939})}\BibitemShut
  {NoStop}%
\end{thebibliography}%
\end{document}